\documentclass {aa}
\usepackage{graphics}

\def\H2 {H$_2$~}

\def\ltsima{$\; \buildrel < \over \sim \;$}
\def\simlt{\lower.5ex\hbox{\ltsima}}    
\def\gtsima{$\; \buildrel > \over \sim \;$}
\def\simgt{\lower.5ex\hbox{\gtsima}}    

\begin{document}

\input psfig.sty

\thesaurus{03(11.01.2; 11.14.1; 11.17.3; 13.09.1)}

\title{Near--infrared imaging of the host galaxies of intermediate redshift 
steep spectrum radio quasars\thanks{Based on observations collected at the 
European Southern Observatory, La Silla, Chile.}}

\author{J. K. Kotilainen\inst{1}, R. Falomo\inst{2}}

\institute{Tuorla Observatory, University of Turku, V\"ais\"al\"antie 20, 
FIN--21500 Piikki\"o, Finland; e-mail: jarkot@deneb.astro.utu.fi
\and 
Osservatorio Astronomico di Padova, vicolo dell'Osservatorio 5, I--35122 
Padova, Italy; e-mail: falomo@pd.astro.it}

\offprints{J.K. Kotilainen}

\date{Received date / Accepted date}

\titlerunning{NIR imaging of SSRQ host galaxies}

\authorrunning{J.K. Kotilainen \& R. Falomo}

\maketitle

\begin{abstract}
We present the results of near--infrared $H$-band (1.65 $\mu$m) imaging of 19 
steep spectrum radio quasars (SSRQ) in the redshift range 0.5 $<$ z $<$ 1.0. 
This sample of SSRQs is matched with our previously studied complete sample 
of 20 flat spectrum radio quasars (FSRQ) with respect to redshift and optical 
and radio luminosity. We are able to clearly detect the host galaxy in 10 
(53 \%) SSRQs and marginally in 6 (32 \%) others, while the host remains 
unresolved in 3 (16 \%) SSRQs. The galaxies hosting the SSRQs are large 
(average bulge scale-length R(e) = 9.0$\pm$1.7 kpc) and luminous (average 
M(H) = -27.2$\pm$1.1). They are, therefore, $\sim$2 mag more luminous than 
the typical galaxy luminosity L* (M*(H) = --25.0$\pm$0.2), and $\sim$1 mag 
more luminous than the brightest cluster galaxies (M(H) = --26.3$\pm$0.3). 
The SSRQ hosts appear to have similar luminosity to those of the FSRQ hosts 
(M(H)$\sim$--27), and they fall between the luminosities of lower redshift 
(M(H) $\sim$--26) and higher redshift (M(H) $\sim$--29) radio-loud quasars. 
The average nucleus--to--galaxy luminosity ratio of SSRQs 
(LN/LG = 3.8$\pm$3.2) is much smaller than that found for the FSRQs 
(LN/LG = 21$\pm$11), in good agreement with the current unified models. We 
confirm for the most luminous SSRQs the trend noted for the FSRQs between the 
host and nuclear luminosity. This trend supports the suggestion based on 
studies of lower redshift AGN, that there is a minimum host galaxy luminosity 
which increases linearly with the quasar luminosity. Finally, FSRQs seem to 
reside in richer environments than SSRQs, as evidenced by a larger number of 
close companion galaxies.


\end{abstract}

\section{Introduction}

Detailed studies of orientation-independent properties of quasars, such as 
their host galaxies and environments, are a fundamental tool to understand 
the quasar phenomenon, and AGN in general. Such studies can crucially test 
models which attempt to unify different observed classes of AGN either 
through evolution (e.g. Ellingson, Yee \& Green 1991) or orientation (e.g. 
Antonucci 1993; Urry \& Padovani 1995). They can shed light on whether 
interactions (as evidenced by close companions or disturbed morphology) are 
important for triggering and fueling of the quasar activity (e.g. 
Smith \& Heckman 1990; Hutchings \& Neff 1992). These studies allow one to 
investigate the possible reciprocal effect of the AGN on the properties and 
evolution of the hosts. Possible cosmological evolution of the AGN population 
(e.g. Small \& Blandford 1992; Silk \& Rees 1998) can be constrained by 
comparing the properties of the hosts and environments of AGN at different 
redshifts. 

The host galaxies of low redshift (z \ltsima 0.3) quasars have been 
thoroughly investigated using both ground-based imaging (e.g. 
McLeod \& Rieke 1995; Taylor et al. 1996; Percival et al. 2000) and the 
Hubble Space Telescope (HST; e.g. Bahcall et al. 1997; Boyce et al. 1998; 
McLure et al. 1999). These investigations have shown that most quasars live 
in galaxies at least as bright as the Shechter function's characteristic 
luminosity L$^*$ (e.g.  Mobasher, Sharples \& Ellis 1993). While most quasar 
host galaxies are brighter than L$^*$, and in many cases comparable to 
brightest cluster member galaxies (BCM; Thuan \& Puschell 1989), some 
undetected or marginally detected hosts may be under-luminous. Recent 
morphological studies of host properties at low $z$ (e.g. Taylor et al. 1996; 
Percival et al. 2000) have concluded that while radio-loud quasars (RLQ) are 
found exclusively in giant elliptical galaxies, radio-quiet quasars (RQQ) 
reside in both elliptical and spiral (disc dominated) galaxies. It has been 
suggested that the morphological type may depend on the power of the quasar, 
with the most luminous quasars found only in spheroidal host galaxies 
(Taylor et al. 1996). Close companions have been found around many quasars 
(e.g. Hutchings 1995; Bahcall et al. 1997; Hutchings et al. 1999), some with 
signs of interaction, but the physical association has been confirmed only in 
some cases through spectroscopic measurements (e.g. Heckman et al. 1984; 
Ellingson et al. 1994; Canalizo \& Stockman 1997).

Our knowledge of the host properties is basically limited to low redshift, 
because of the increasing difficulty of both resolving the quasar and the 
rapid cosmological dimming ($\alpha$ (1+z)$^4$) of the host in contrast to 
the nucleus. At intermediate redshift (0.5 $<$ z $<$ 1), quasar hosts have 
been little investigated (e.g. Carballo et al. 1998; R\"onnback et al. 1996). 
Recently, we studied in the near-infrared (NIR) the host galaxies of a 
complete sample of 20 flat spectrum radio quasars (FSRQ) (most of which are 
at 0.5 $<$ z $<$ 1) (Kotilainen, Falomo \& Scarpa 1998; hereafter KFS98). We 
were able to resolve the host galaxy in a substantial fraction ($\sim$50 \%) 
of them and found these highly polarized quasars to reside in galaxies of 
M(H)$\sim$--27.  

While most FSRQs are characterized by rapid variability, high and variable 
polarization, high brightness temperatures, core--dominated radio emission 
and confirmed or suspected superluminal motion (e.g. Impey \& Tapia 1990; 
Quirrenbach et al. 1992; Padovani \& Urry 1992; Vermeulen \& Cohen 1994), 
very few of these characteristics are shared by the more common steep 
spectrum radio quasars (SSRQ). This difference is usually explained in the 
unified model in terms of synchrotron radiation strongly relativistically 
beamed close to our line--of--sight in FSRQs, while SSRQs are viewed further 
away from the beaming axis (Blandford \& Rees 1978). 

We present here deep high spatial resolution ($\sim$0\farcs3 px$^{-1}$, 
$\leq$ 1$''$ FWHM seeing) NIR $H$-band (1.65 $\mu$m) imaging study of the 
host galaxies of a sample of SSRQs in the redshift range 0.5 $<$ z $<$ 1. We 
thus explore the host galaxies in the rest-frame 0.8 -- 1.1 $\mu$m, offering 
many advantages over the optical wavelengths, including larger host/nucleus 
contrast, negligible scattered light from the quasar, small K-correction and 
insignificant extinction (see KFS98 for details). For practically all the 
quasars in the sample we present the first high quality NIR imaging 
observations, and the first detection of the host galaxy. 

\begin{table*}
\begin{center}
 \caption{General properties of the SSRQ sample, journal of observations, and 
NIR photometry\label{tab:gen}}
 \begin{tabular}{lllllllllll}
\hline
Name & Other name & z & $\alpha$(6-11 cm) & M(B) & log L(11 cm) & Date & I/S$^a$ & T(exp) & FWHM & 6$''$ ap.\\
     &            &   &                   &      & (erg/sec) &      & & (min.) & (arcsec)   & (mag)\\
PKS 0056--00 & PHL 923 & 0.717 & 0.46 & --25.8 & 44.58 & 31/12/97 & I & 42 & 0.9 & 14.09\\
            &         &       &      &       &       & 03/01/98 & I & 20 & 1.0 & 14.24\\
PKS 0159--11 & 3C 57 & 0.669 & 0.61 & --26.6 & 44.53 & 30/12/97 & I & 48 & 0.9 & 13.08\\
PKS 0349--14 & 3C 95 & 0.614 & 1.12 & --26.4 & 44.30 & 02/01/98 & I & 60 & 0.8 & 12.89\\
PKS 0413--21 &       & 0.807 & 0.54 & --25.3 & 44.69 & 03/01/98 & I & 60 & 1.1 & 14.87\\
PKS 0414--06 & 3C 110 & 0.773 & 0.83 & --27.3 & 44.12 & 31/01/99 & S & 36 & 0.7 & 14.55\\
PKS 0454--22 &       & 0.534 & 0.97 & --26.4 & 44.12 & 31/01/99 & S & 36 & 0.7 & 14.02\\
PKS 0518+16 & 3C 138 & 0.759 & 0.60 & --24.1 & 45.15 & 30/12/97 & I & 48 & 1.0 & 14.76\\
            &        &       &       &      &       & 31/12/97 & I & 60 & 1.1 & 14.65\\
PKS 0710+11 & 3C 175 & 0.768 & 1.03 & --26.4 & 44.47 & 02/01/98 & I & 42 & 1.0 & 13.09\\
            &        &       &       &      &       & 03/01/98 & I & 64 & 1.0 & 13.14\\
PKS 0825--20 &        & 0.822 & 0.76 & --25.8 & 44.78 & 30/12/97 & I & 64 & 1.0 & 14.29\\
            &        &       &      &       &       & 02/01/98 & I & 54 & 0.9 & 14.25\\
PKS 0838+13 & 3C 207 & 0.684 & 0.35 & --24.6 & 44.51 & 01/02/99 & S & 54 & 0.9 & 15.92\\
PKS 0855--19 &        & 0.660 & 0.04 & --24.4 & 44.11 & 31/01/99 & S & 48 & 0.7 & 15.44\\
PKS 0903--57 &        & 0.695 & 0.45 & --24.2 & 44.48 & 30/12/97 & I & 48 & 1.2 & 15.46\\
            &        &       &       &      &       & 31/12/97 & I & 60 & 1.0 & 15.60 \\
PKS 0959--443 &       & 0.837 & 0.08 & --26.8 & 44.42 & 31/01/99 & S & 48 & 0.8 & 14.68\\
PKS 1046--409 &       & 0.620 & 0.47 & --24.9 & 44.29 & 03/01/98 & I & 64 & 0.9 & 13.88\\
PKS 1116--46 &        & 0.713 & 0.35 & --26.0 & 44.51 & 31/01/99 & S & 48 & 0.6 & 15.90\\
PKS 1136--13 &        & 0.554 & 0.59 & --26.3 & 44.46 & 31/12/97 & I & 42 & 0.9 & 13.35\\
PKS 1237--10 &        & 0.750 & 0.00 & --25.4 & 44.54 & 31/01/99 & S & 52 & 0.7 & 16.04\\
PKS 1327--21 &        & 0.528 & 0.48 & --25.7 & 44.10 & 31/01/99 & S & 44 & 0.7 & 15.07\\
PKS 1335--06 &        & 0.625 & 0.97 & --25.2 & 44.44 & 31/01/99 & S & 18 & 0.8 & 16.40\\
            &        &       &      &       &        & 01/02/99 & S & 36 & 1.0 & 16.16\\
\hline
\end{tabular}
\end{center}
$^a$: I = IRAC2, S = SOFI.\\
\end{table*} 

\begin{figure}
\psfig{file=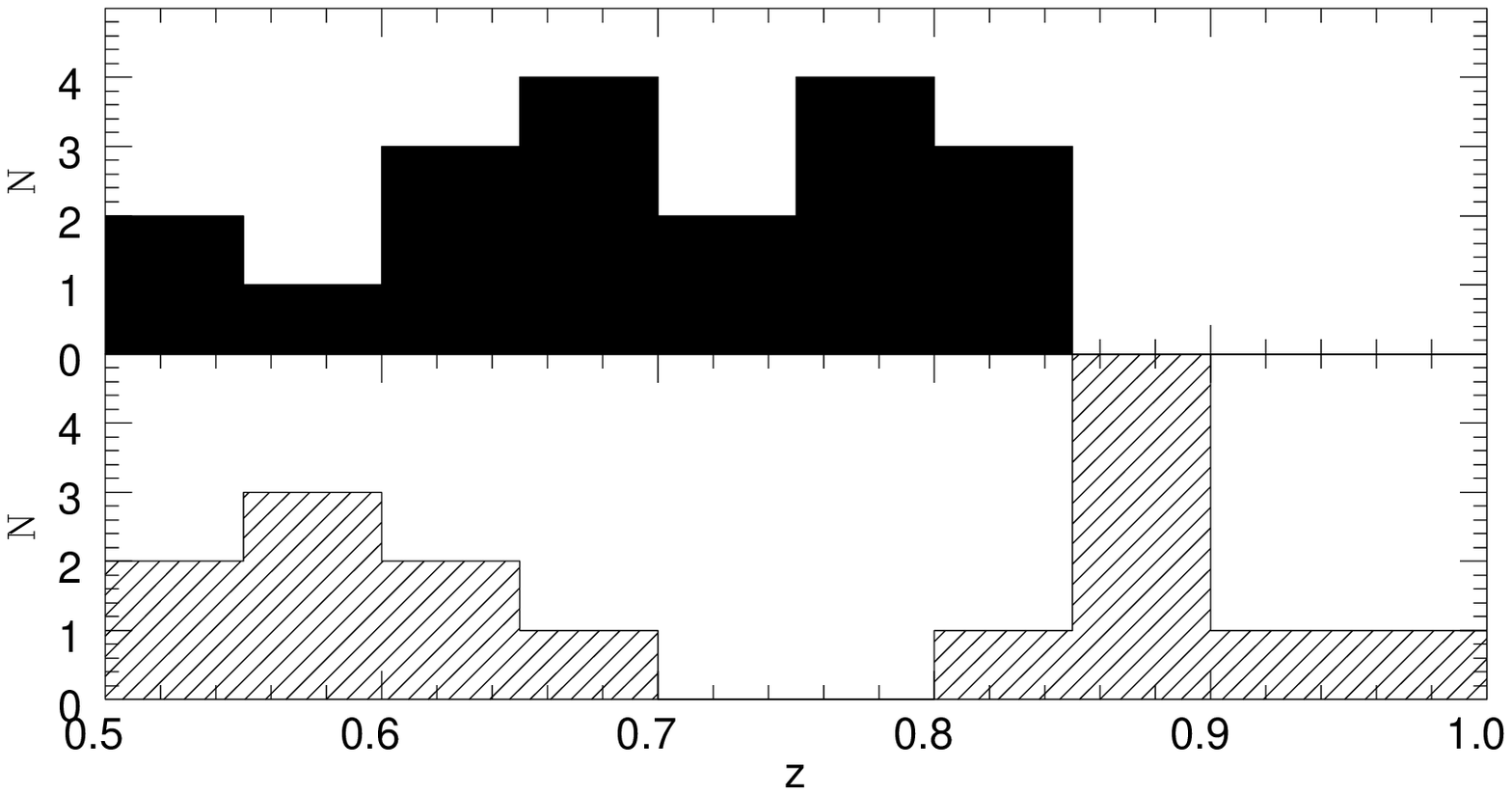,width=9cm,height=4.5cm}
\psfig{file=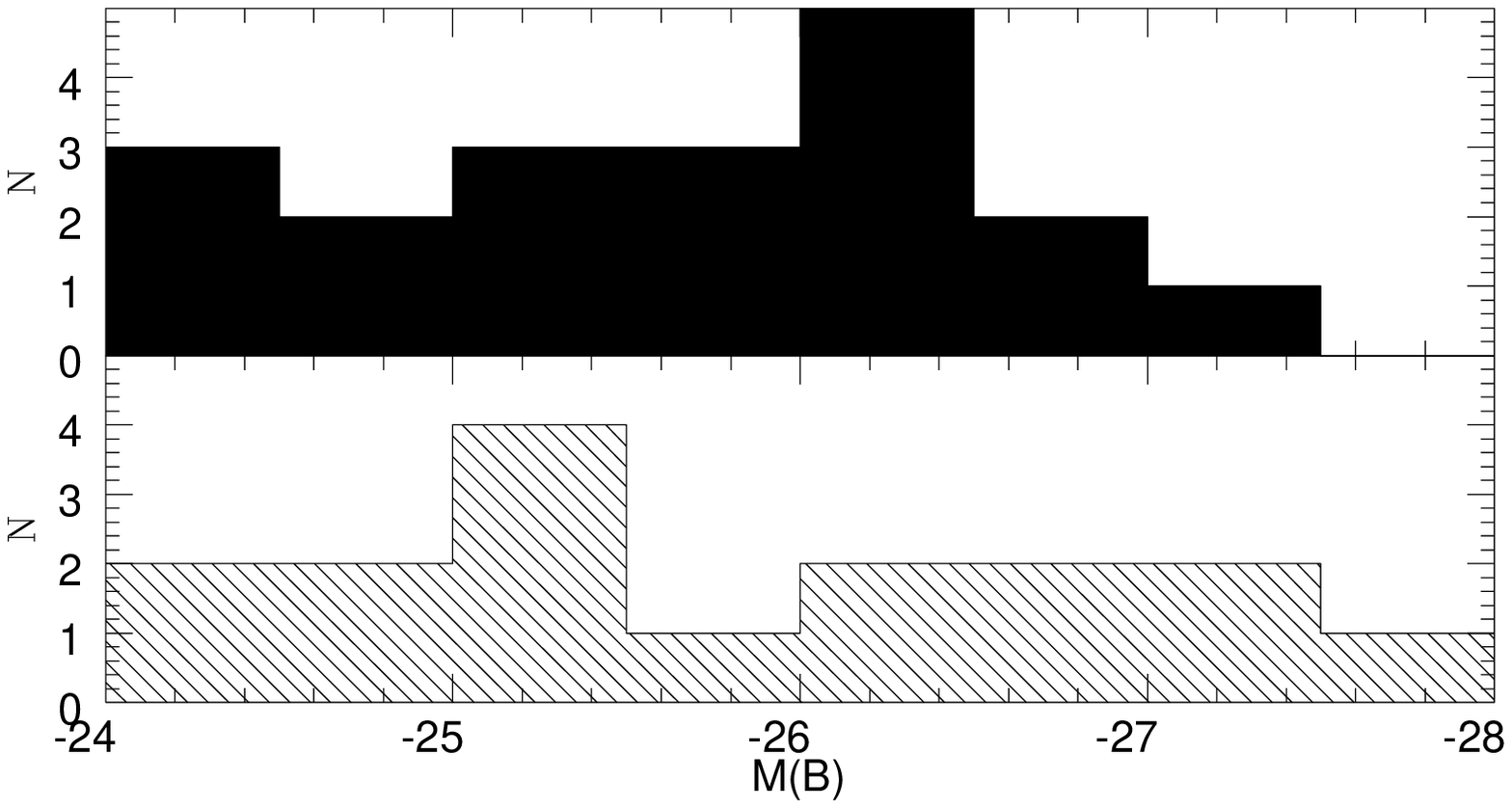,width=9cm,height=4.5cm}
\psfig{file=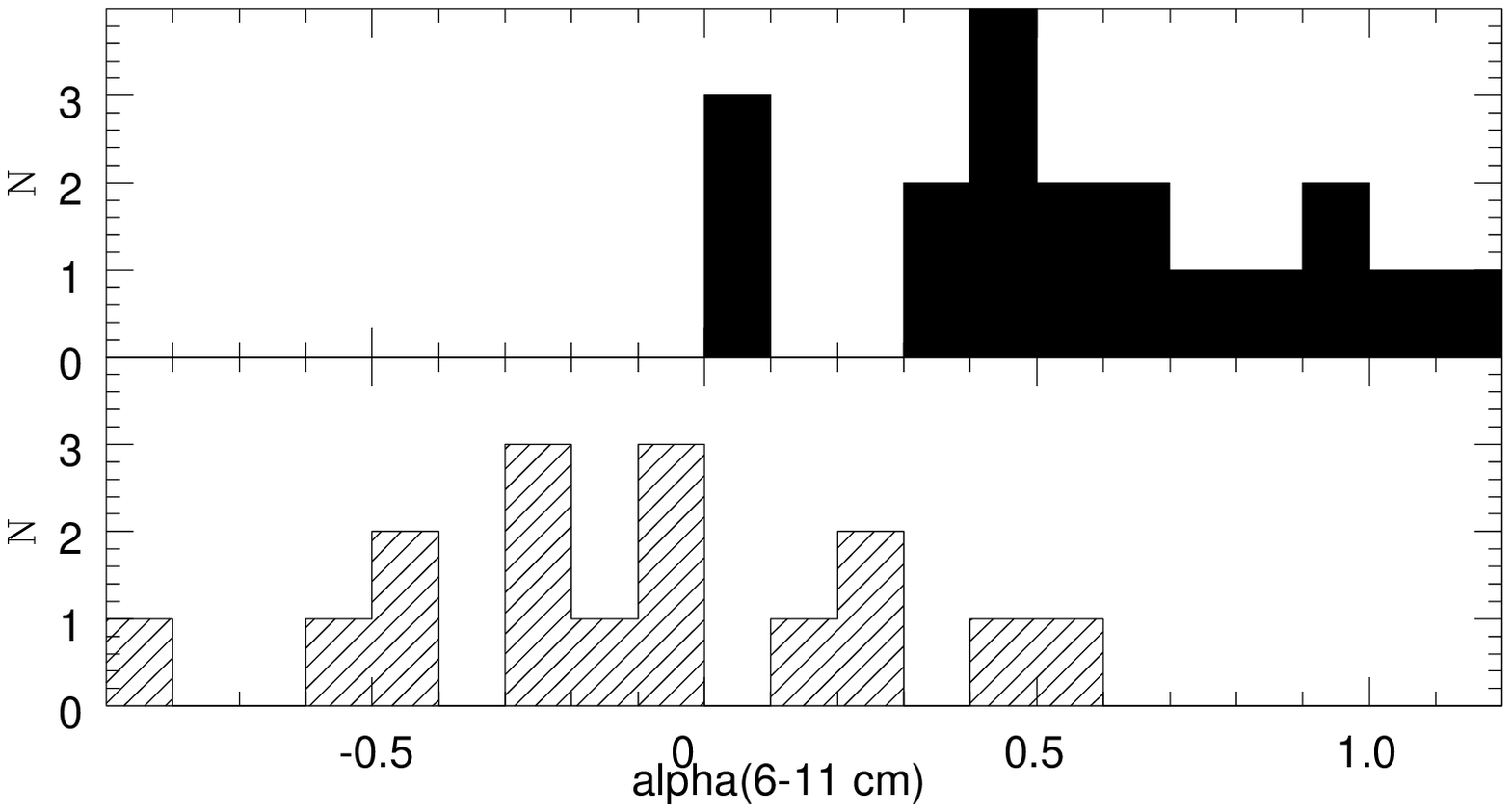,width=9cm,height=4.5cm}
\psfig{file=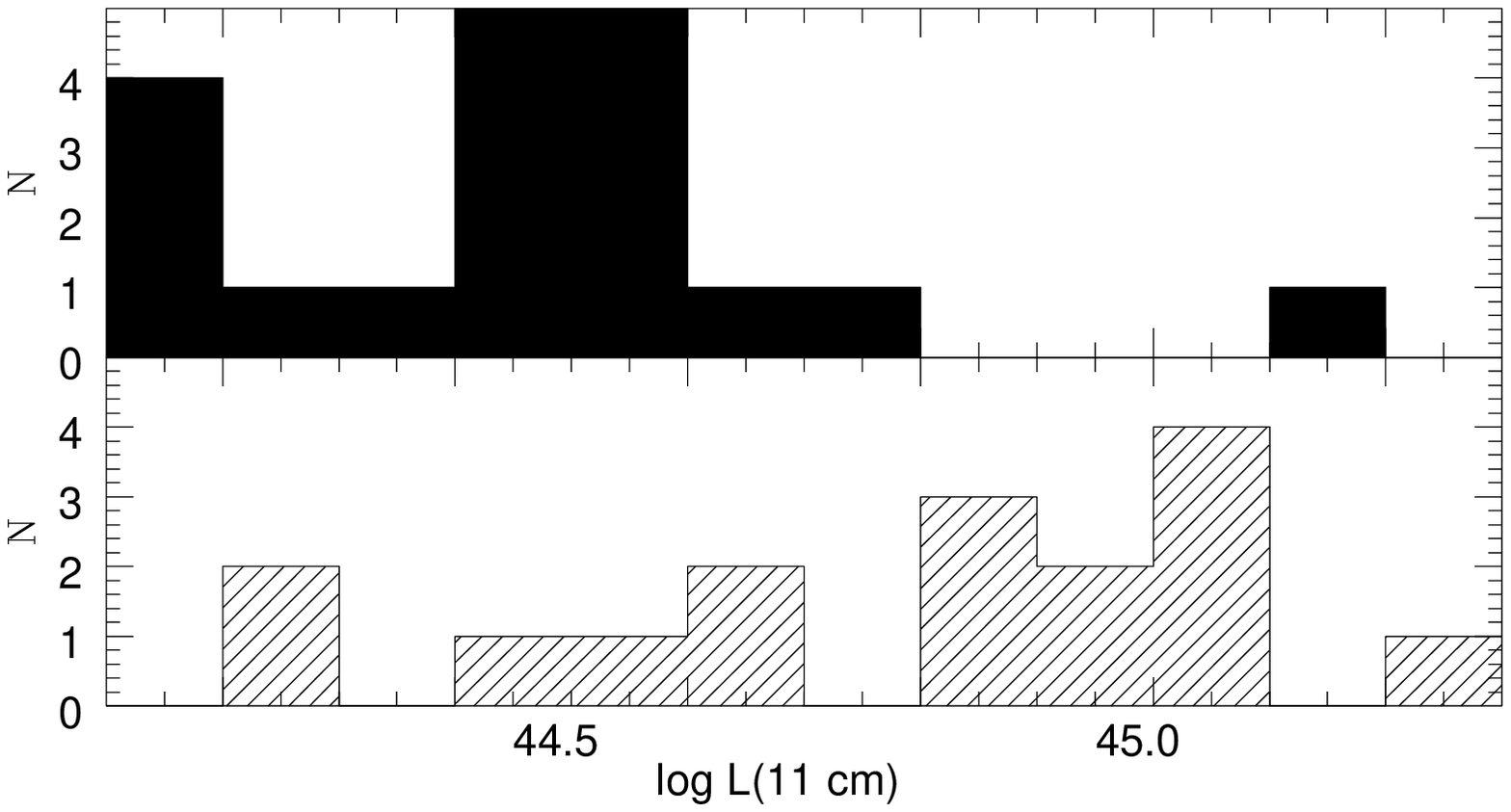,width=9cm,height=4.5cm}
\caption{Comparison, from top to bottom, of the redshift, M(B), 
$\alpha$(6--11 cm) and L(11 cm) distributions of the SSRQ (filled histogram) 
and FSRQ (hatched histogram) samples.}
\label{fig:compar}
\end{figure}

\begin{table}
\begin{center}
 \caption{Average properties of the SSRQ and FSRQ samples.\label{tab:aver}}
 \begin{tabular}{lllll}
\hline
sample & z & M(B) & $\alpha$(6--11 cm) & log L(11 cm)\\
SSRQ   & 0.691$\pm$0.092 & --25.7$\pm$0.9 & 0.56$\pm$0.32 & 44.45$\pm$0.25\\
FSRQ   & 0.751$\pm$0.166 & --25.8$\pm$1.0 & 0.02$\pm$0.37 & 44.78$\pm$0.29\\
\hline
\end{tabular}
\end{center}
\end{table} 

The SSRQ sample was chosen from the V\'eron-Cetty \& V\'eron (1998) AGN catalogue 
to match the FSRQ sample (KFS98) in terms of redshift and optical and radio 
luminosity distribution, while keeping the two samples separated regarding 
the radio spectral index. The selection criteria applied were: 
0.5 $<$ z $<$ 1.0, $\delta$ $<$ +20\degr, M$_B$ $<$ -24.0, 
$\alpha$(6-11 cm) $>$ 0, and 
L(11 cm) $>$ 10$^{44}$ erg sec$^{-1}$. This selection yielded a total of 54 
SSRQs. Out of these, 33 were observable during the observing runs. A total of 
19 SSRQs out of these were observed, and their general properties are given 
in Table~\ref{tab:gen}. The properties of the SSRQ and FSRQ samples are 
compared in Fig.~\ref{fig:compar} and Table~\ref{tab:aver}. The M$_B$ 
distribution is very similar for the two samples. The FSRQ sample has more 
high redshift quasars, but the average value is in agreement with the SSRQs. 
The samples are reasonably well separated in $\alpha$, although some overlap 
exists.  Finally, there is a marked difference in the L(11 cm) distribution, 
in the sense that the FSRQ distribution extends to much higher radio 
luminosities than the SSRQ sample. This is due to the unfortunate 
circumstance that there were not enough very radio-luminous SSRQs available 
at the time of the observations.  This caveat should be kept in mind in the 
following discussion.

The outline of the paper is as follows. In section 2, we briefly describe the 
observations, data reduction and the method of the analysis. Our results 
concerning the host galaxies and the nuclear component are given in section 3 
and compared with the results for FSRQs. In section 4, we analyze the close 
environment of SSRQs and FSRQs and, finally, in section 5 summary and main 
conclusions of this study are drawn. In the Appendix, we give comments for 
individual quasars. Throughout this paper, 
H$_{0}$ = 50 km s$^{-1}$ Mpc$^{-1}$ and q$_{0}$ = 0 are used. 

\section{Observations, data reduction and modeling of the luminosity profiles}

The observations were carried out during two observing runs. The first was at 
the 2.2m telescope of European Southern Observatory (ESO), La Silla, Chile, 
between 30 December 1997 and 3 January 1998, using the 256x256 px IRAC2 NIR 
camera (Moorwood et al. 1992) with pixel scale 0\farcs278 px$^{-1}$, giving a 
field of view of 1.2 arcmin$^2$. The second run was at the ESO 3.5m 
New Technology Telescope (NTT) on the nights of 31 January and 
1 February 1999, using the 1024x1024 px SOFI NIR camera 
(Lidman \& Cuby 1998), with pixel scale 0\farcs292 px$^{-1}$, giving a field 
of view of 5.0 arcmin$^2$. Details of the observations and NIR photometry are 
given in Table~\ref{tab:gen}. The observations, data reduction and modeling 
of the luminosity profiles were performed following the procedure described 
in KFS98. Briefly, targets were always kept in the field by shifting them in 
a grid across the array, with typical offsets of 30$''$. Individual exposures 
of 60 sec duration were coadded to achieve the final integration time 
reported in Table~\ref{tab:gen}.

Data reduction, performed using {\sc{IRAF}}\footnote{IRAF is distributed by 
the National Optical Astronomy Observatories, which are operated by the 
Association of Universities for Research in Astronomy, Inc., under 
cooperative agreement with the National Science Foundation.}, consisted of 
correction for bad pixels by interpolating across neighboring pixels, sky 
subtraction using a median averaged, scaled sky frame from all observations 
of a given target, flat-fielding using illuminated dome exposures, and 
combination of all images of the same target, using field stars or the 
centroid of the light distribution of the quasar as a reference point. 
Standard stars from Landolt (1992) were used for photometric calibration, for 
which we estimate an accuracy of $\sim$0.1 mag, based on comparison of 
multiple observations at different nights with IRAC2 as listed in 
Table~\ref{tab:gen}. K--correction was applied to the host galaxy magnitudes 
following the method of Poggianti (1997). The applied K--correction for each 
quasar is reported in Table~\ref{tab:prop}. This correction, insignificant at 
low redshift (m(H) = 0.03 at z = 0.2), has to be taken into account at 
moderate redshifts (m(H)$\sim$0.14 at our median redshift of z = 0.65). No 
K--correction was applied to the nuclear component. 

After masking all the regions around the target contaminated by companions, 
azimuthally averaged radial luminosity profiles were extracted for each 
quasar and field stars out to the background noise level, corresponding to a 
surface brightness of $\mu$(H) = 23--24 mag arcsec$^{-2}$.  In the small 
field of view IRAC2 images, bright stars were present in the field of only a 
few quasars. For most quasars, the core of the PSF was derived from faint 
field stars, while the wing was derived from separate observation of a nearby 
bright star or extrapolated using a suitable Moffat (1969) function obtained 
from fitting bright stars in other frames of similar seeing during the same 
night. In the much larger field--of--view SOFI images, all quasars had bright 
stars in the field, enabling a good PSF determination.

The luminosity profiles were fitted into a point source (PSF) and a galaxy 
(de Vaucouleurs law, convolved with the PSF) component by an iterative 
least--squares fit to the observed profile. For these moderately high 
redshift quasars, data quality does not allow one to discriminate between 
elliptical and disk morphologies for the host galaxy, and the elliptical 
model was assumed, as demonstrated to be the case for low redshift RLQs (e.g. 
Taylor et al. 1996; Bahcall et al. 1997). For quasars with no host galaxy 
detection, we determined an upper limit to the brightness 
of the host galaxy by adding simulated ``host galaxies'' of various 
brightness to the observed profile until the simulated host became detectable 
within the errors of the luminosity profile. For a detailed discussion of 
potential error sources in this fitting procedure, see KFS98. We estimate the 
uncertainty of the derived host galaxy magnitudes to be $\sim$$\pm$0.3 mag 
for the clearly detected hosts. For the marginally detected hosts, we can 
only assess a lower limit to the error margin as $\geq$$\pm$0.5. 

For the clearly detected hosts, there is expectedly a good agreement between 
the total magnitude of the quasars derived from aperture photometry 
(Table~\ref{tab:gen}, column 11) and that resulting from the profile fit 
(Table~\ref{tab:prop}, columns 5 and 6), with average magnitude difference of 
--0.03$\pm$0.12.

\section{Results and discussion}

\begin{figure*}
\psfig{file=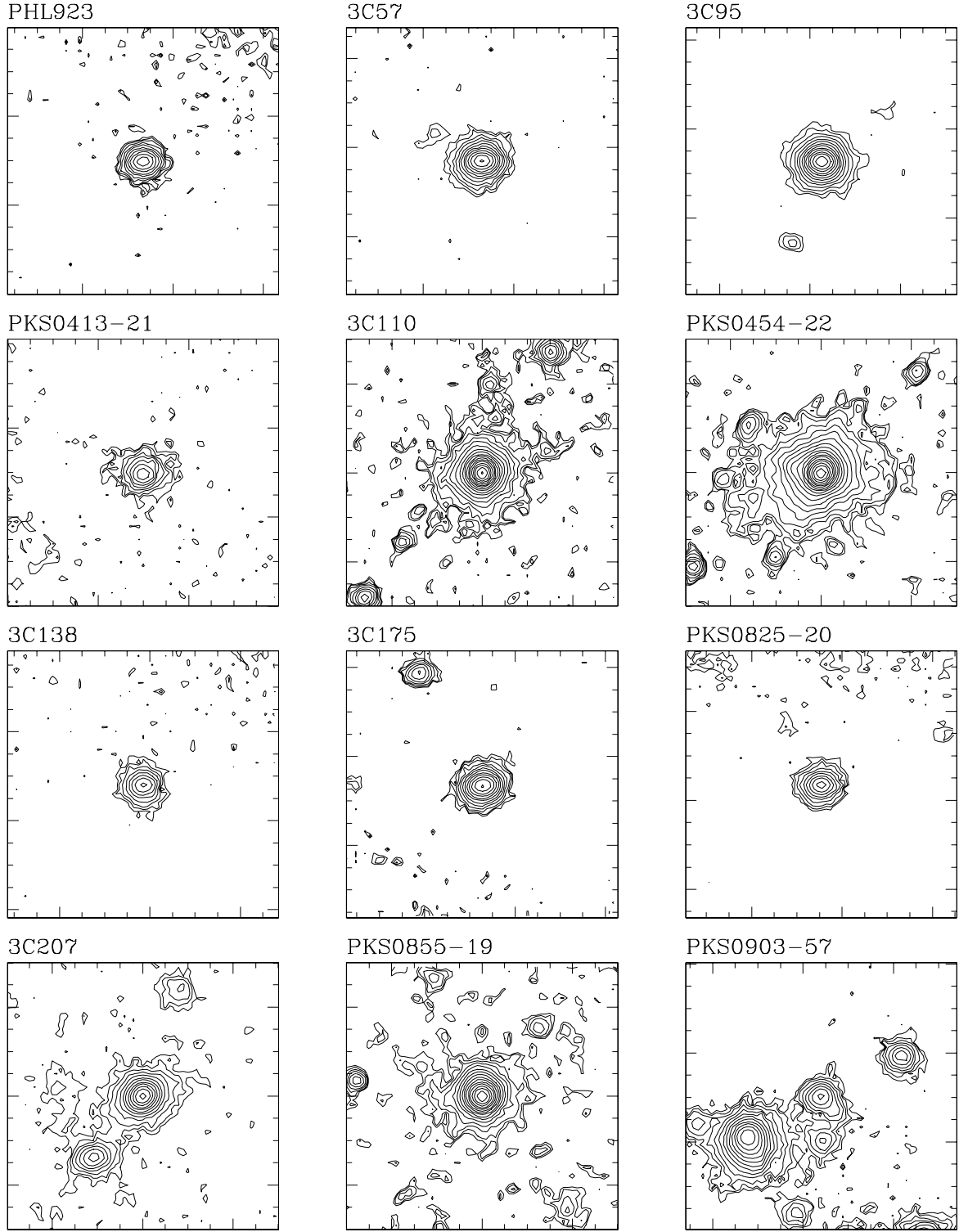,width=18cm,height=21.1cm}
\caption{Gaussian ($\sigma$ = 1 px) smoothed contour plots of the SSRQs in 
the $H$-band. The SSRQ is the object in the center of the frame. The full 
size of the image is 60 px across (corresponding to 16\farcs7 and 17\farcs5 
for IRAC2 and SOFI data, respectively). The contours are separated by 0.5 mag 
intervals. North is up and east to the left.}
\label{fig:cont}
\end{figure*}

\begin{figure*}
\psfig{file=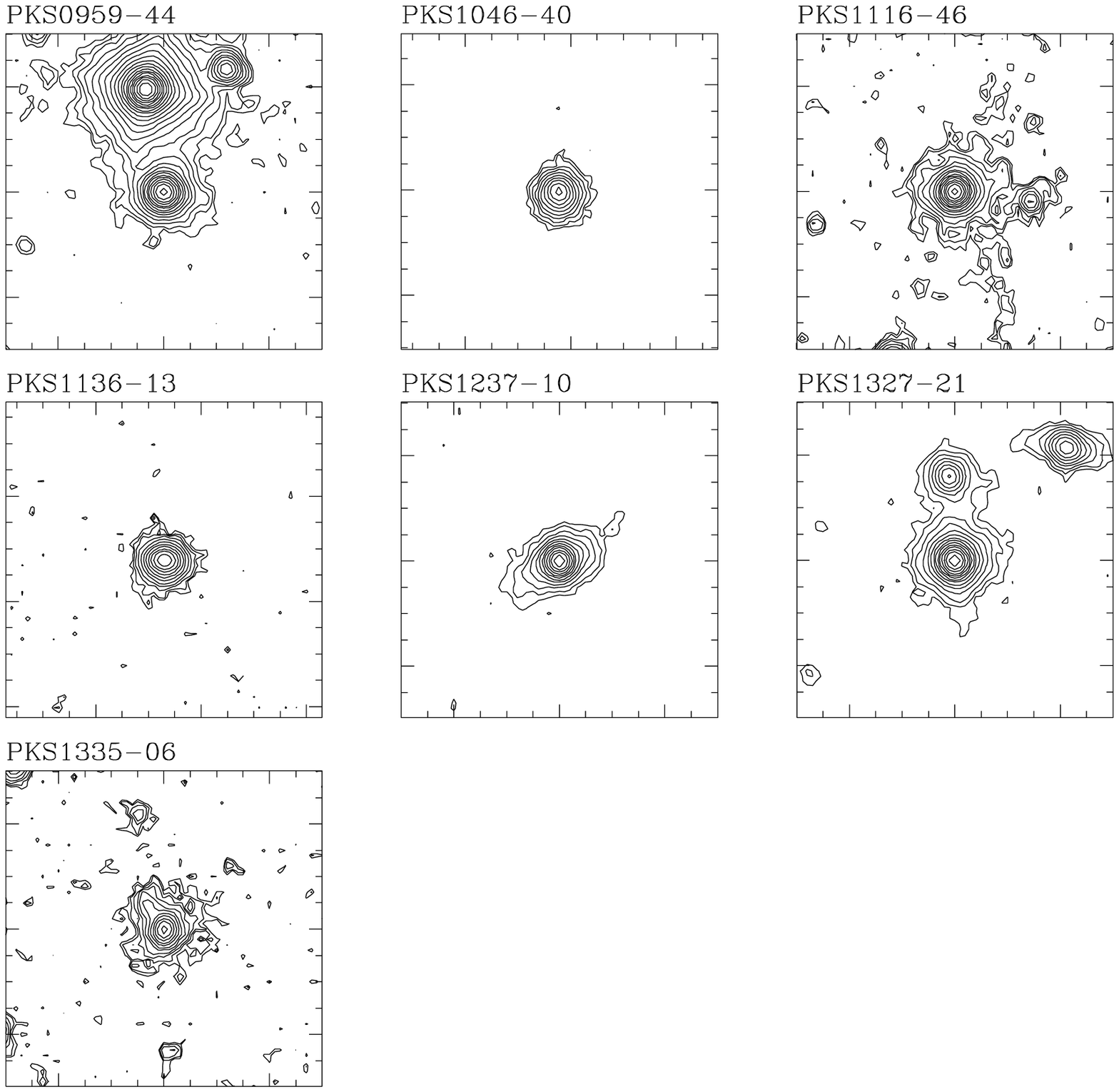,width=18cm,height=16.5cm}
{\bf Fig. 2.} continued.
\end{figure*}

\begin{figure*}
\psfig{file=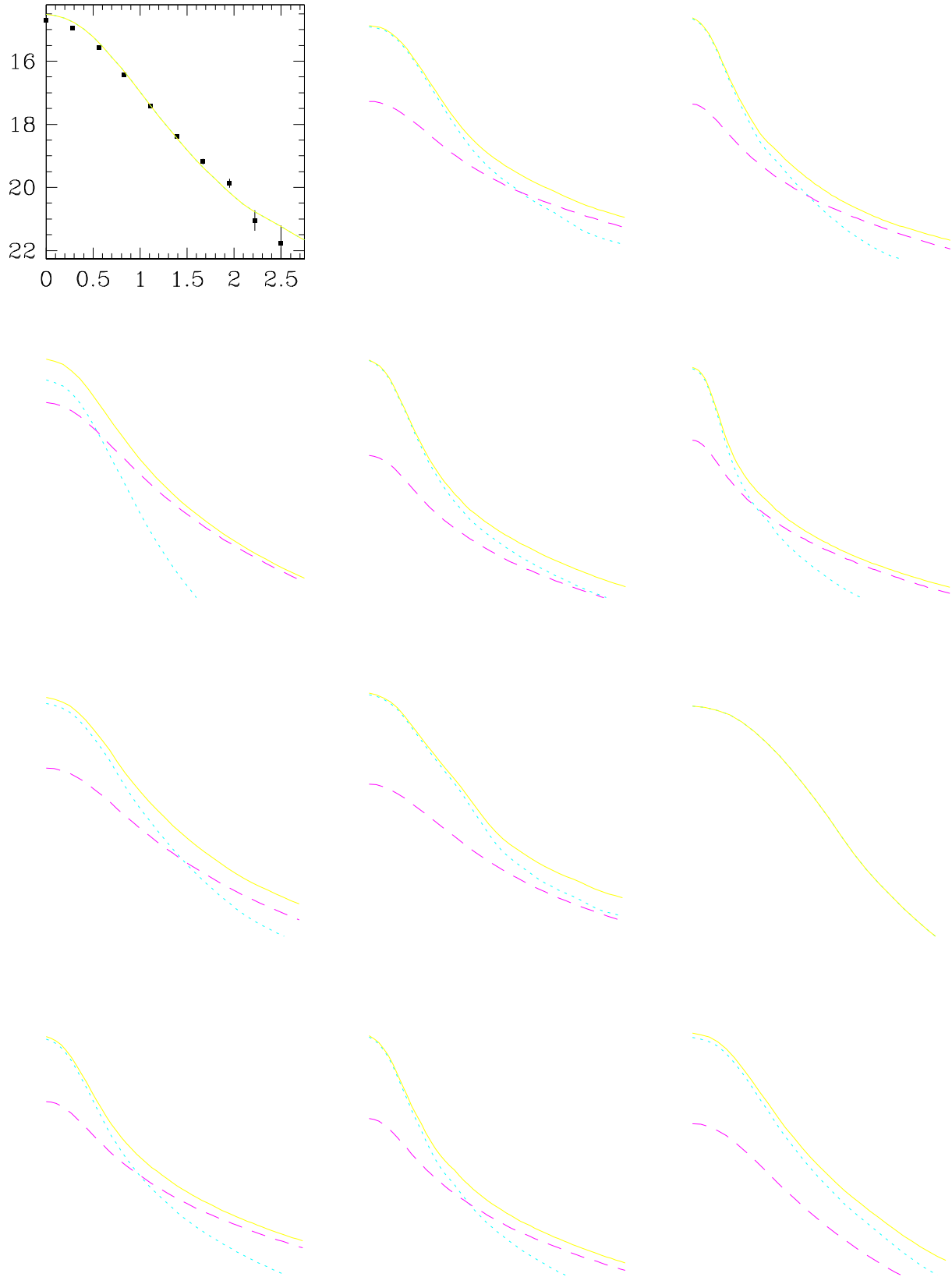,width=18cm,height=23.5cm}
\caption{The observed radial luminosity profiles of each SSRQ (filled 
squares), superimposed to the fitted  model consisting of the PSF 
(short-dashed line), and the de Vaucouleurs bulge (long-dashed line). The 
solid line shows the total model fit.}
\label{fig:prof}
\end{figure*}

\begin{figure*}
\psfig{file=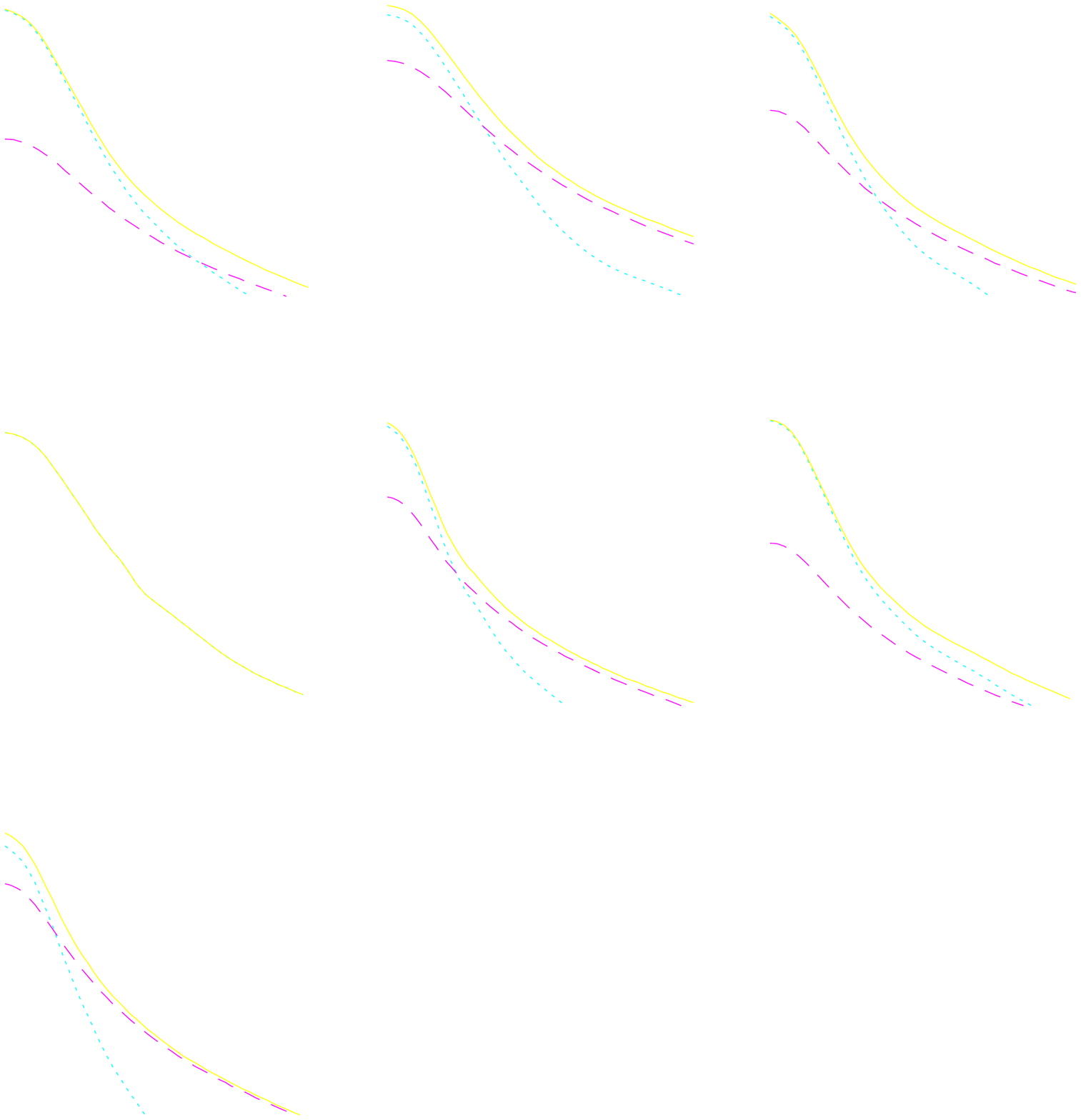,width=18cm,height=18cm}
{\bf Fig. 3.} continued.
\end{figure*}

\begin{table*}
\begin{center}
 \caption{Properties of the host galaxies.\label{tab:prop}}
 \begin{tabular}{llllllllll}
\hline
Name & z & K--corr. & r(e)/R(e) & m(H) & m(H) & L(n)/L(g) & M(H) & M(H) & Note$^a$\\
 & & mag & arcsec/kpc & nucleus & host & & nucleus & host & \\
PKS 0056--00 & 0.717 & 0.17 & & 14.1 & $>$16.7 & $>$12 & --29.7 & $>$--27.0 & U\\
PKS 0159--11 & 0.669 & 0.15 & 0.94/8.8 & 16.1 & 18.2 & 7.0 & --27.6 & --25.4 & M\\
PKS 0349--14 & 0.614 & 0.14 & 1.20/11.2 & 13.0 & 15.1 & 7.0 & --30.6 & --28.5 & R\\
PKS 0413--21 & 0.807 & 0.20 & 0.95/9.6 & 15.8 & 15.1 & 0.51 & --28.3 & --29.1 & R\\
PKS 0414--06 & 0.773 & 0.19 & 0.81/8.0 & 14.8 & 17.4 & 11.4 & --29.3 & --26.6 & R\\
PKS 0454--22 & 0.534 & 0.12 & 1.45/12.1 & 14.4 & 15.9 & 4.1 & --28.6 & --27.1 & R\\
PKS 0518+16 & 0.759 & 0.18 & 0.81/8.0 & 17.9 & 18.6 & 2.0 & --26.1 & --25.4 & M\\
PKS 0710+11 & 0.768 & 0.19 & 0.81/8.0 & 13.2 & 15.3 & 7.0 & --30.8 & --28.7 & M\\
PKS 0825--20 & 0.822 & 0.20 & & 18.8 & $>$19.1 & $>$1.4 & --25.4 & $>$--25.0 & U\\
PKS 0838+13 & 0.684 & 0.15 & 0.95/8.9 & 16.3 & 17.5 & 3.0 & --27.4 & --26.2 & R\\
PKS 0855--19 & 0.660 & 0.14 & 1.00/9.3 & 15.7 & 17.5 & 5.1 & --27.9 & --26.1 & R\\
PKS 0903--57 & 0.695 & 0.16 & 0.45/4.3 & 14.9 & 16.5 & 4.3 & --28.9 & --27.3 & M\\
PKS 0959--44 & 0.837 & 0.21 & 1.08/11.1 & 14.9 & 16.9 & 6.1 & --29.4 & --27.4 & M\\
PKS 1046--40 & 0.620 & 0.14 & 0.70/6.3 & 14.4 & 14.6 & 1.3 & --29.0 & --28.8 & R\\
PKS 1116--46 & 0.713 & 0.17 & 0.95/9.1 & 16.3 & 17.4 & 2.8 & --27.5 & --26.4 & R\\
PKS 1136--13 & 0.554 & 0.12 & & 13.5 & $>$15.6 & $>$7.6 & --29.7 & $>$--27.5 & U\\
PKS 1237--10 & 0.750 & 0.18 & 0.80/7.8 & 16.5 & 17.4 & 2.4 & --27.5 & --26.6 & R\\
PKS 1327--21 & 0.528 & 0.12 & 0.96/8.0 & 15.2 & 17.6 & 8.6 & --27.8 & --25.4 & M\\
PKS 1335--06 & 0.625 & 0.14 & 0.80/7.2 & 17.4 & 17.2 & 0.81 & --26.1 & --26.3 & R\\
\hline
\end{tabular}
\end{center}
$^a$ R = resolved; M = marginally resolved; U = unresolved.\\
\end{table*} 

\begin{table*}
\begin{center}
 \caption{Comparison of the average host galaxy properties with other 
samples.}
\label{tab:comp}
 \begin{tabular}{lcrllll}
\hline
Sample & filter & N & $<z>$ & $<M_B>$ & $<M_H(nuc)>$ & $<M_H(host)>$ $^b$\\
L$^*$ Mobasher et al. (1993) & K & 136 & 0.077$\pm$0.030 & & & --25.0$\pm$0.2 \\
                            &   &     &     &     & &                \\
BCM Thuan \& Puschell (1989) & H & 84 & 0.074$\pm$0.026 & & & --26.3$\pm$0.3 \\
                            &   &     &     &     & &                \\
RLQ McLeod \& Rieke (1994a) & H & 22 & 0.103$\pm$0.029 &  & --25.1$\pm$0.5 & --24.9$\pm$0.6 \\
RLQ McLeod \& Rieke (1994b) & H & 23 & 0.196$\pm$0.047 &  & --26.5$\pm$0.9 & --25.7$\pm$0.6 \\
RLQ McLure et al. (1999) & R & 6 & 0.219$\pm$0.029 & & --26.8$\pm$0.7 & --26.4$\pm$0.3\\
RLQ Bahcall et al. (1997) & V & 6 & 0.220$\pm$0.047 & --25.5$\pm$0.9  & & --26.1$\pm$0.5 \\
RLQ Taylor et al. (1996) & K & 13 & 0.236$\pm$0.046 & --24.5$\pm$0.8 & --27.1$\pm$0.8 & --26.3$\pm$0.7 \\
RLQ V\'eron-Cetty \& Woltjer (1990) & I & 20 & 0.343$\pm$0.094 & --25.2$\pm$0.5 & & --26.3$\pm$0.5 \\
RLQ Hooper et al. (1997) & R & 6 & 0.465$\pm$0.032 &  & --26.8$\pm$0.4 & --26.2$\pm$0.4 \\
RLQ Lehnert et al. (1992) & K & 6 & 2.342$\pm$0.319 &  & --30.5$\pm$1.0 & --28.8$\pm$1.1 \\
                            &   &     &     &     & &                \\
RQQ Taylor et al. (1996) & K & 19 & 0.157$\pm$0.062 & --23.8$\pm$0.6 & --26.1$\pm$0.9 & --25.7$\pm$0.7 \\
RQQ McLure et al. (1999) & R & 9 & 0.169$\pm$0.033 & & --25.5$\pm$1.9 & --25.8$\pm$0.5\\
RQQ Bahcall et al. (1997) & V & 14 & 0.183$\pm$0.046 & --24.9$\pm$0.5  & &    --25.1$\pm$0.6\\
RQQ Percival et al. (2000) & K & 14 & 0.357$\pm$0.057 & --25.6$\pm$0.8 & --28.0$\pm$0.8 & --24.9$\pm$0.5\\
RQQ Hooper et al. (1997) & R & 10 & 0.433$\pm$0.032 & & --25.8$\pm$0.8 &     --25.6$\pm$0.5\\
                            &   &     &     &     & &                \\
FSRQ/R+M$^a$ (KFS98; 0.5$<$z$<$1.0) & H & 9 & 0.671$\pm$0.157 & --26.2$\pm$1.1 & --29.7$\pm$0.8 & --26.7$\pm$1.2 \\
FSRQ/R$^a$ (KFS98; 0.5$<$z$<$1.0) & H & 4 & 0.673$\pm$0.141 & --25.9$\pm$1.3  & --30.2$\pm$0.7 & --27.8$\pm$0.3 \\
                    &   &     &     &     & &                \\
SSRQ/R+M$^a$ & H & 16 & 0.690$\pm$0.088 & --25.6$\pm$1.0 & --28.3$\pm$1.3 & --27.0$\pm$1.2 \\
SSRQ/R$^a$ & H & 10 & 0.683$\pm$0.077 & --25.6$\pm$0.9 & --28.2$\pm$1.2 & --27.2$\pm$1.1 \\
\hline
\end{tabular}
\end{center}
$^a$ R = resolved; M = marginally resolved.\\
$^b$ Transformation of magnitudes to $H$-band done assuming V--H = 3.0, R--H = 
2.5 and H--K = 0.2 galaxy colours, and H$_0$ = 50 km s$^{-1}$ Mpc$^{-1}$ ~and 
q$_0$ = 0.
\end{table*} 

In Fig.~\ref{fig:cont}, we show the $H$-band contour plots of all the SSRQs, 
after smoothing the images with a Gaussian filter of $\sigma$ = 1 px. In 
Fig.~\ref{fig:prof} we show the radial luminosity profiles of each quasar, 
with the best--fit decomposition overlaid. We detect the host galaxy clearly 
for 10 (53 \%) SSRQs and marginally for 6 (32 \%) more. The host remains 
unresolved for 3 (16 \%) SSRQs. We summarize the 
best--fit model parameters of the profile fitting and the derived properties 
of the host galaxies in Table~\ref{tab:prop}, where column (1) gives the name 
of the SSRQ, (2) the redshift, (3) the applied K--correction, (4) the 
scale-length of the host galaxy, (5) -- (6) the apparent magnitude of the 
nucleus and the host, (7) the nucleus/host luminosity ratio, (8) -- (9) the 
absolute magnitude of the nucleus and the host, and column (10) note about 
the detection of the host.

Table~\ref{tab:comp} presents a comparison of the $H$-band absolute 
magnitudes of the SSRQ hosts with relevant samples from previous studies in 
the literature, for which we report the average values after correcting the 
published values for colour term and to our cosmology. In the Appendix, we 
compare our NIR photometry with the few previous studies, and discuss in more 
detail individual quasars, including comparison with the few previous 
optical/NIR determinations of the host galaxies. 

\subsection{The host galaxies}

\begin{figure}
\psfig{file=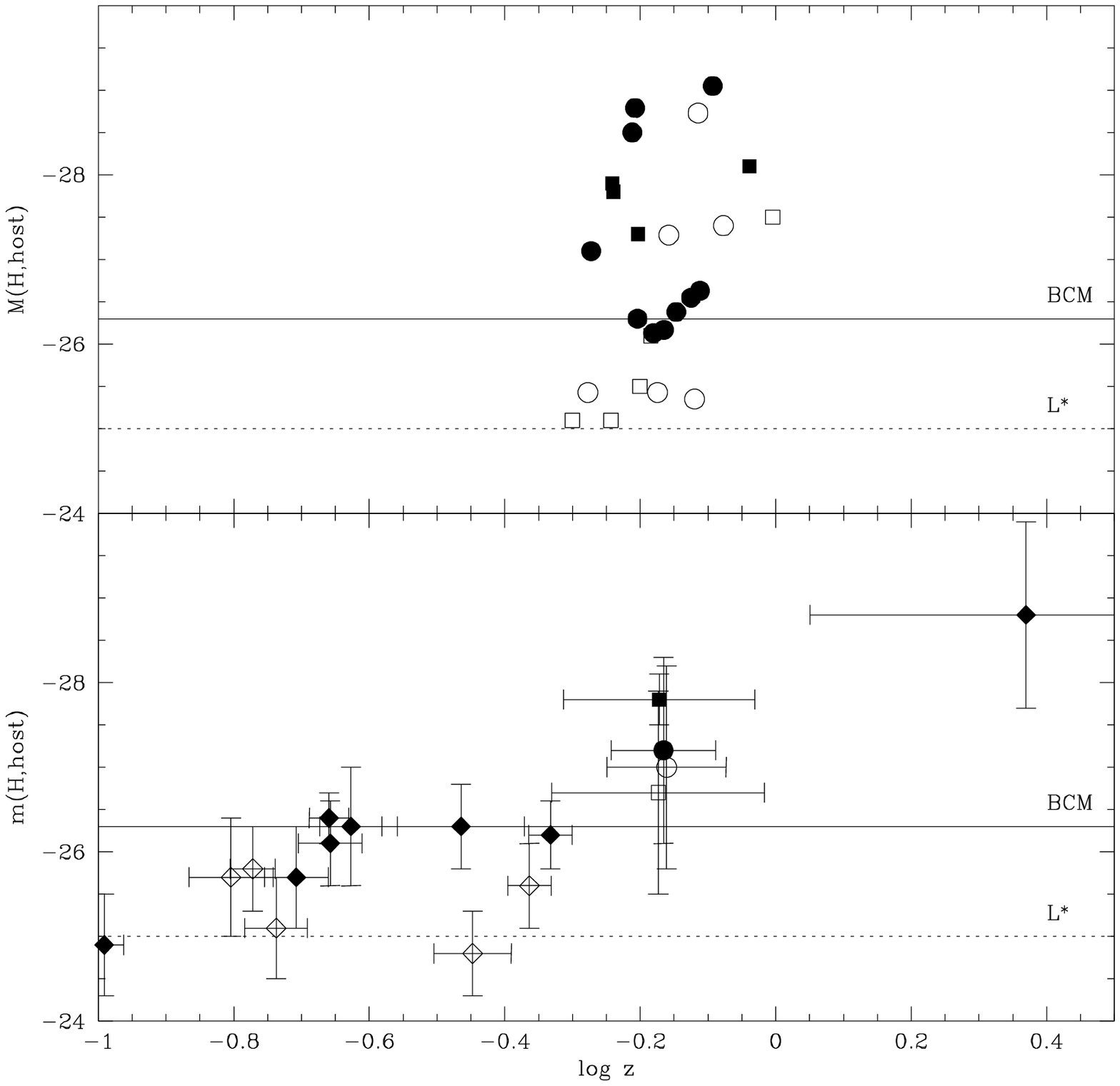,width=9cm,height=9cm}
\caption{{\bf Upper panel:} The absolute $H$-band magnitude of the host 
galaxies of SSRQs and FSRQs vs. redshift. Resolved SSRQs are marked as filled 
circles, and marginally resolved SSRQs as open circles. FSRQs from KFS98 are 
marked as filled (resolved) and open (marginally resolved) squares. The 
average luminosities of L* galaxies (M(H)$\sim$--25.0; Mobasher et al. 1993) 
and brightest cluster member galaxies (BCM; M(H)$\sim$--26.3; 
Thuan \& Puschell 1989) are indicated as dashed and solid lines,respectively. 
{\bf Lower panel:} As in the upper panel, except for the mean values of the 
SSRQs and FSRQs in comparison with RLQ (filled diamonds) and RQQ (open 
diamonds) samples from literature.}
\label{fig:Mhz}
\end{figure}

In Fig.~\ref{fig:Mhz} (upper panel) we investigate the location of the SSRQ 
and FSRQ hosts in the $H$-band absolute magnitude vs. redshift diagram. In 
Fig.~\ref{fig:Mhz} (lower panel) we show the same diagram for the mean value 
of various samples of RLQ and RQQ hosts from the literature. The average 
$H$-band absolute magnitude of the resolved SSRQ host galaxies is 
M(H) = --27.2$\pm$1.1 and the average bulge scale-length 
R(e) = 9.0$\pm$1.7 kpc, while the values after adding the marginally resolved 
hosts are M(H) = --27.0$\pm$1.2 and R(e) = 8.6$\pm$1.9 kpc. The SSRQ hosts 
are therefore large (all have R(e) $>$ 3 kpc, the empirical upper boundary 
found for normal local ellipticals by Capaccioli, Caon \& D'Onofrio 1992), 
and very luminous, much brighter than the luminosity of an L$^*$ galaxy, 
which has M(H) = --25.0$\pm$0.3 (Mobasher et al. 1993). It is therefore 
evident that the SSRQ hosts are preferentially selected from the 
high--luminosity tail of the galaxy luminosity function (the derived upper 
limits for the unresolved hosts are also consistent with this; see 
Table~\ref{tab:prop}). Indeed, we find no case of an SSRQ host with 
M(H) $>$ --25, indicating that for some reason these quasars cannot be hosted 
by a galaxy with L $<$ L$^*$, similarly to what was found by 
Taylor et al. (1996) for low redshift RLQs. Although there is a large spread, 
the SSRQ hosts are on average of similar luminosity to those of FSRQ hosts 
(KFS98). 

There is a suggestion of a positive correlation of host luminosity with 
redshift (Fig.~\ref{fig:Mhz}, lower panel) as the FSRQ and SSRQ hosts fall 
between the luminosities of lower redshift (M(H) $\sim$--26) and higher 
redshift (M(H) $\sim$--29) RLQs. This is consistent with what is expected 
from passive stellar evolution models for elliptical galaxies (e.g. 
Bressan et al. 1994; Fukugita, Shimasaku \& Ichikawa 1995), and the evolution 
of galaxies in clusters (Ellingson et al. 1991), although the scatter is 
quite large. On the other hand, it is not consistent with models of 
hierarchical galaxy formation (e.g. Kaufmann \& H\"ahnelt 2000), which 
predict $\sim$L$^*$ hosts at z = 2 -- 3 that afterwards undergo mergers to 
become present-day giant elliptical hosts. The detected correlation, 
therefore, suggests that major merging has already happened at z = 2 - 3. 
Also, this is consistent with the discovery of ellipticals with old stellar 
populations even at z$\sim$2 (e.g. Spinrad et al. 1997; 
Stiavelli et al. 1999). However, this scenario seems not to be valid for 
RQQs, which show no or little evolution with redshift, as evidenced by the 
non-detection of high redshift RQQ hosts (Lowenthal et al. 1995). 

On the other hand, the detected correlation can be interpreted as evidence 
for a relationship between optical and radio luminosities. 
Ledlow \& Owen (1996) found such a correlation for radio galaxies and in the 
unified model (e.g. Urry \& Padovani 1995), where RLQs and radio galaxies are 
identical objects seen from different viewing angles, this relationship is 
expected also to exist between the radio luminosities and host galaxy 
absolute magnitudes for RLQs at different redshifts. 

Most of the available comparison data from literature are for low and 
intermediate redshift RLQs. These samples span a moderately large range in 
redshift from z$\sim$0.1 up to z$\sim$0.7, which is the average redshift of 
the SSRQ sample. The average host galaxy magnitudes for the various samples 
are given in Table~\ref{tab:comp} in order of increasing average redshift. 
Considering all these samples together gives average host magnitude of 
M(H) = --26.2$\pm$0.7. As can be seen from Fig.~\ref{fig:Mhz}, there is no 
significant difference among the average values of these low z samples.  The 
average M(H) = --27.0$\pm$1.2 for the SSRQ hosts is therefore 
$\sim$1 magnitude brighter than the luminosity of low z RLQ which suggests 
evolution in the host brightness with redshift, and/or a relationship of the 
host luminosity with the nuclear luminosity (see section 3.2).

It is worth to compare these data also with those for samples of low redshift 
RQQs extracted from the literature. The average host galaxy magnitudes for 
the various samples are given in Table~\ref{tab:comp}. Considering all these 
RQQ samples together yields an average host magnitude of 
M(H) = --25.4$\pm$0.4, $\sim$1 magnitude fainter than comparable redshift RLQ 
hosts and $\sim$1.5 mag fainter than SSRQ hosts at intermediate redshift, 
indicating that RLQs and RQQs inhabit different types of galaxies. However, 
this difference may also be due to the above mentioned correlation between 
optical and radio luminosities (Ledlow \& Owen 1996), in the sense that RQQ 
surveys select objects according to the optical luminosity, while RLQ surveys 
select objects with high radio luminosity and, therefore, a bright host 
galaxy. 

Lehnert et al. (1992, 1999) have reported spatially resolved structures 
around RLQs at z$\sim$1.5 and z$\sim$2.3 that, if interpreted as host 
galaxies, would correspond to extremely luminous galaxies with average 
M(H) = --29.1$\pm$1.1, $\sim$2 mag brighter than the FSRQs and SSRQs at 
z$\sim$0.7. However, within the scatter involved in these numbers, our 
results appear to be consistent with those of Lehnert et al., 
for the trend between the nuclear and host galaxy luminosities (see 
section 3.2), and is supporting evidence for the existence of a real upturn 
in the host luminosity occurring between z$\sim$0.5 and z$\sim$2, leading 
from L$\geq$L* hosts at low redshift to the host galaxies of high redshift 
quasars that are several magnitudes brighter than L* (see 
Fig.~\ref{fig:Mhz}). While this type of change is consistent with evolution 
of the stellar population in the elliptical hosts (as argued for high 
redshift RGs by Lilly \& Longair 1984), or being intrinsic AGN luminosity 
effect (as argued for high redshift RGs by Eales et al. 1997), there are many 
caveats in this comparison, most notably differences in the intrinsic quasar 
luminosity of the various samples. In addition, optical and NIR imaging by 
Lowenthal et al. (1995) failed to detect extended emission in a sample of six 
RQQs at z$\sim$2.3. Their upper limits (average m(H) $>$ 18.8$\pm$0.4 and 
M(H) $<$ -28.4$\pm$0.6) indicate that the RQQ hosts at high redshift must be 
$\leq$3 mag brighter than L* and $\geq$1 mag fainter than those found in the 
Lehnert et al. (1992) sample of RLQs at similar redshift and with roughly 
similar nuclear luminosity, again suggesting that RLQs and RQQs are different 
types of objects. 

\begin{figure}
\psfig{file=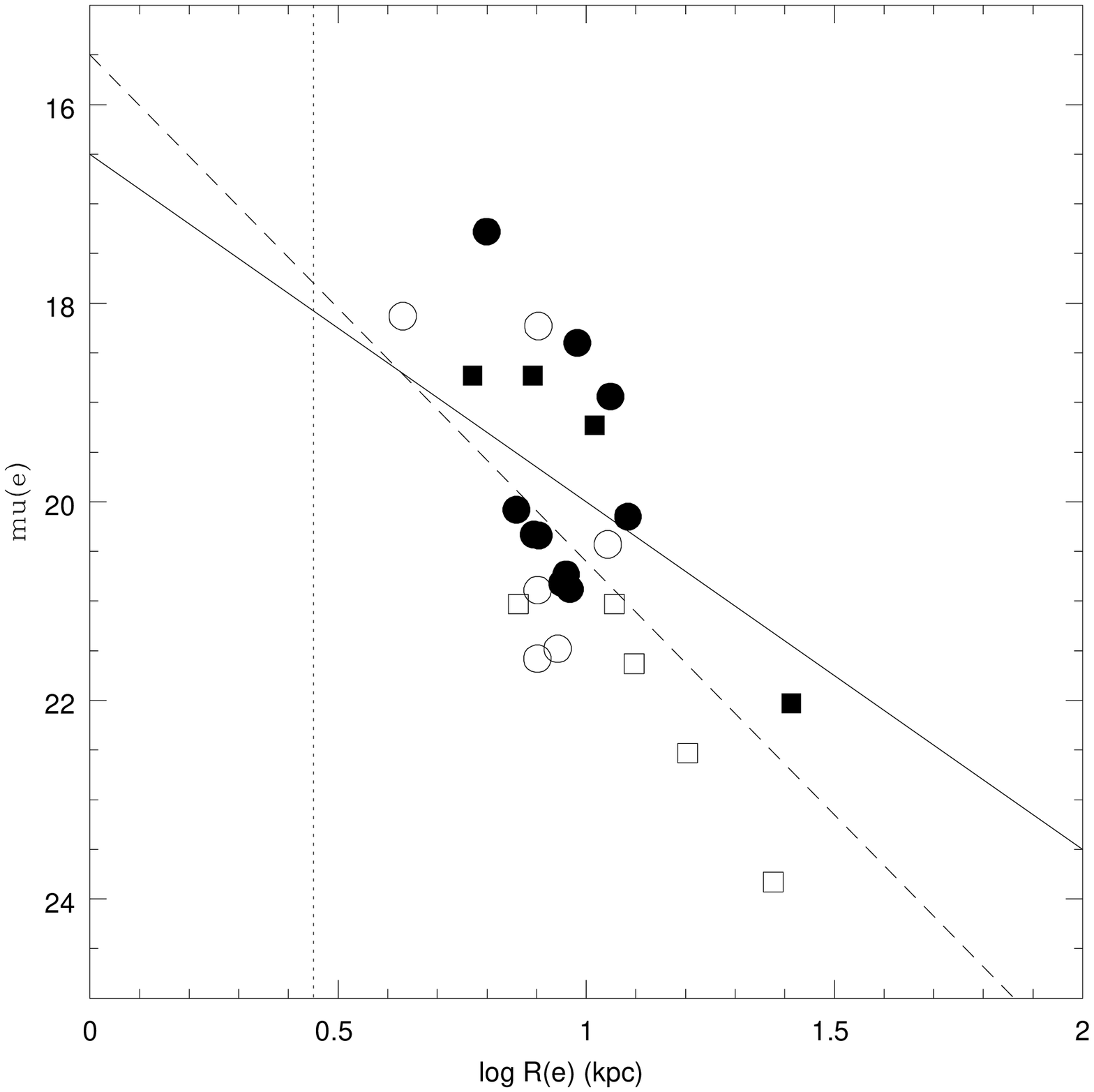,width=9cm,height=9cm}
\caption {Kormendy relation for SSRQ (filled circles) and FSRQ (open circles) 
host galaxies. The solid line shows the relation for z$\sim$0.8 3CR RGs 
(McLure \& Dunlop 2000), the long-dashed line constant galaxy luminosity, and 
the short-dashed vertical line the dividing line between normal and giant 
ellipticals (Capaccioli et al. 1992).}
\label{fig:mur}
\end{figure}

It is well established that elliptical galaxies form families of homologous 
systems with characteristic parameters  R(e), $\mu$(e) and velocity 
dispersion $\sigma$. These are commonly represented in the Fundamental Plane 
(e.g. Djorgovski \& Davis 1987). These relations have been claimed to be 
related to the morphological and dynamical structure of the galaxies, and to 
their formation process. We have investigated the properties of the SSRQ and 
FSRQ hosts in the projected Fundamental Plane (F-P) concerning the effective 
surface brightness $\mu$(e) and the effective radius R(e). Surface brightness 
$\mu$(e) was corrected for Galactic extinction, K--correction and for the 
(1+z)$^4$ cosmological dimming. Fig.~\ref{fig:mur} shows the correlation 
between $\mu$(e) and log R(e) for the SSRQ and FSRQ hosts. It can be seen 
that the behavior of the SSRQ and FSRQ hosts are similar. Both follow the 
Kormendy (1977) relation for giant massive ellipticals (e.g. Capaccioli et 
al. 1992). No host galaxy is in the (scatter) area at log R(e) $<$ 0.5 kpc. 
This confirms that the SSRQ and FSRQ hosts are exclusively drawn from the 
population of giant ellipticals. 

\subsection{The nuclear component}

\begin{figure}
\psfig{file=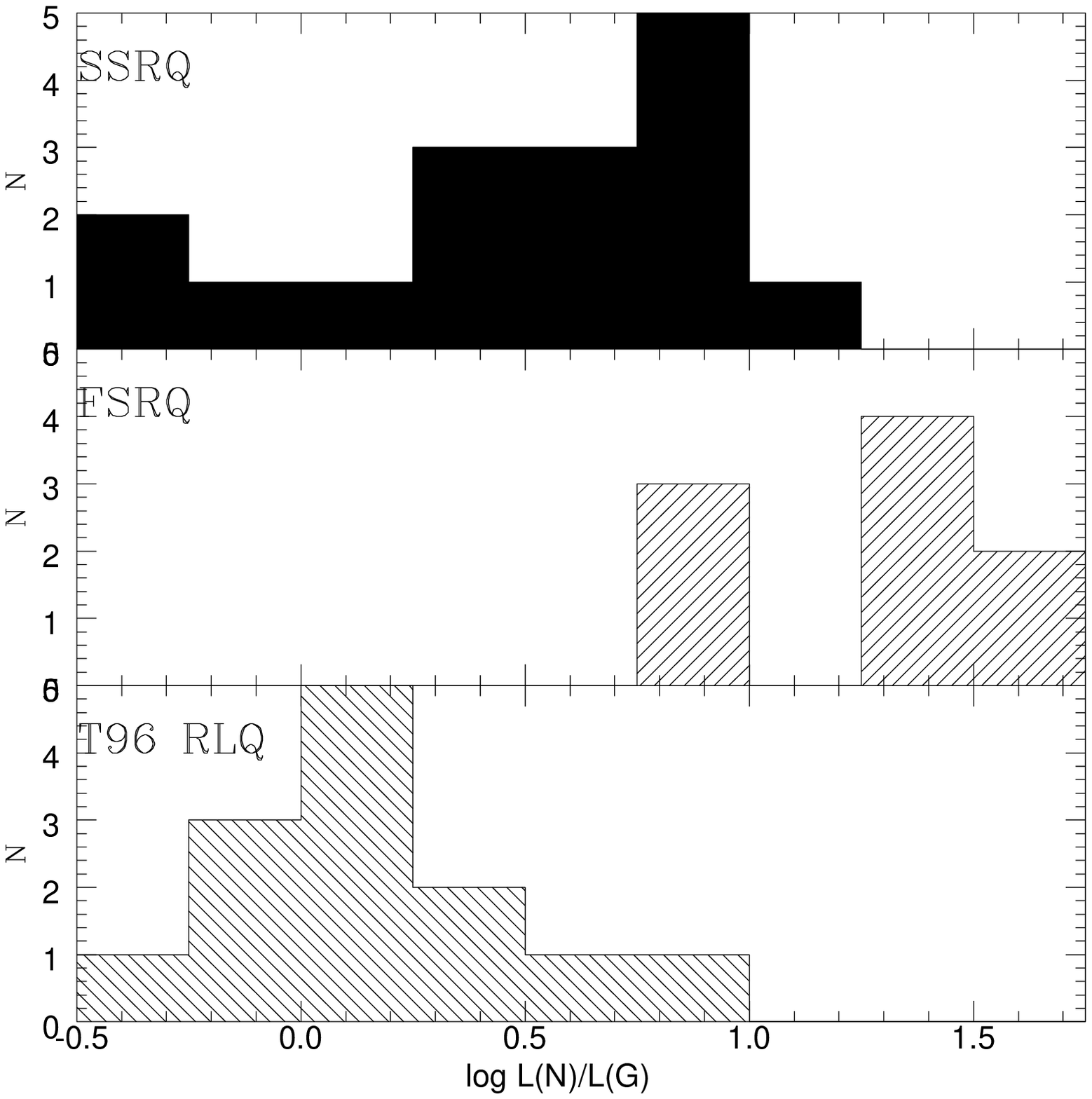,width=9cm,height=9cm}
 \caption{Histogram of the nucleus/host luminosity ratio for the SSRQs (top 
panel), FSRQs (middle panel; from KFS98) and low redshift RLQs (bottom panel; 
from Taylor et al. 1996).}
\label{fig:lnlghist}
\end{figure}

The average absolute magnitude of the nuclear component for all SSRQs is 
M(H) = --28.2$\pm$1.2, that is $\sim$1.5 mag brighter than low redshift RLQ 
nuclei (average M(H) = --26.8$\pm$0.9) and $\sim$2 mag brighter than low 
redshift RQQ nuclei (average M(H) = --26.4$\pm$1.0). This is likely due to a 
selection effect present in the original samples induced by the different 
average redshift. On the other hand, the SSRQ nuclei are $\sim$1.5 mag 
fainter than the nuclei of FSRQs (KFS98; M(H) = --29.7$\pm$0.8) in the 
similar redshift range. This difference is even more evident when considering 
the nucleus/galaxy (LN/LG) luminosity ratio, shown in 
Fig.~\ref{fig:lnlghist}. The average LN/LG ratios are 3.8$\pm$3.2 and 
21$\pm$11 for the SSRQs and FSRQs, respectively. The large range in the 
luminosity ratio can be due to differences in the intrinsic nuclear or host 
luminosity, or a difference in the beaming factor from one object to another. 
Whereas the majority of the FSRQs have LN/LG $>$ 10 in the $H$-band, only one 
SSRQ is above this limit, and the distribution of the LN/LG ratio for the 
SSRQs is similar to that of the low redshift RLQs (e.g. Taylor et al. 1996).

From Fig.~\ref{fig:Mhz} it appears that the host galaxies of the various 
samples are not dramatically different in intrinsic luminosity, especially if 
stellar evolution in the elliptical host galaxies is taken into account. 
Therefore, Fig.~\ref{fig:lnlghist} clearly indicates that SSRQs exhibit a 
nuclear component which is systematically brighter than that of lower 
redshift RLQs and RQQs, but fainter than that of the FSRQs. This is 
consistent with the beaming model with larger Doppler amplification factor 
for FSRQs than SSRQs, that makes the observed differences understandable. 

\begin{figure}
\psfig{file=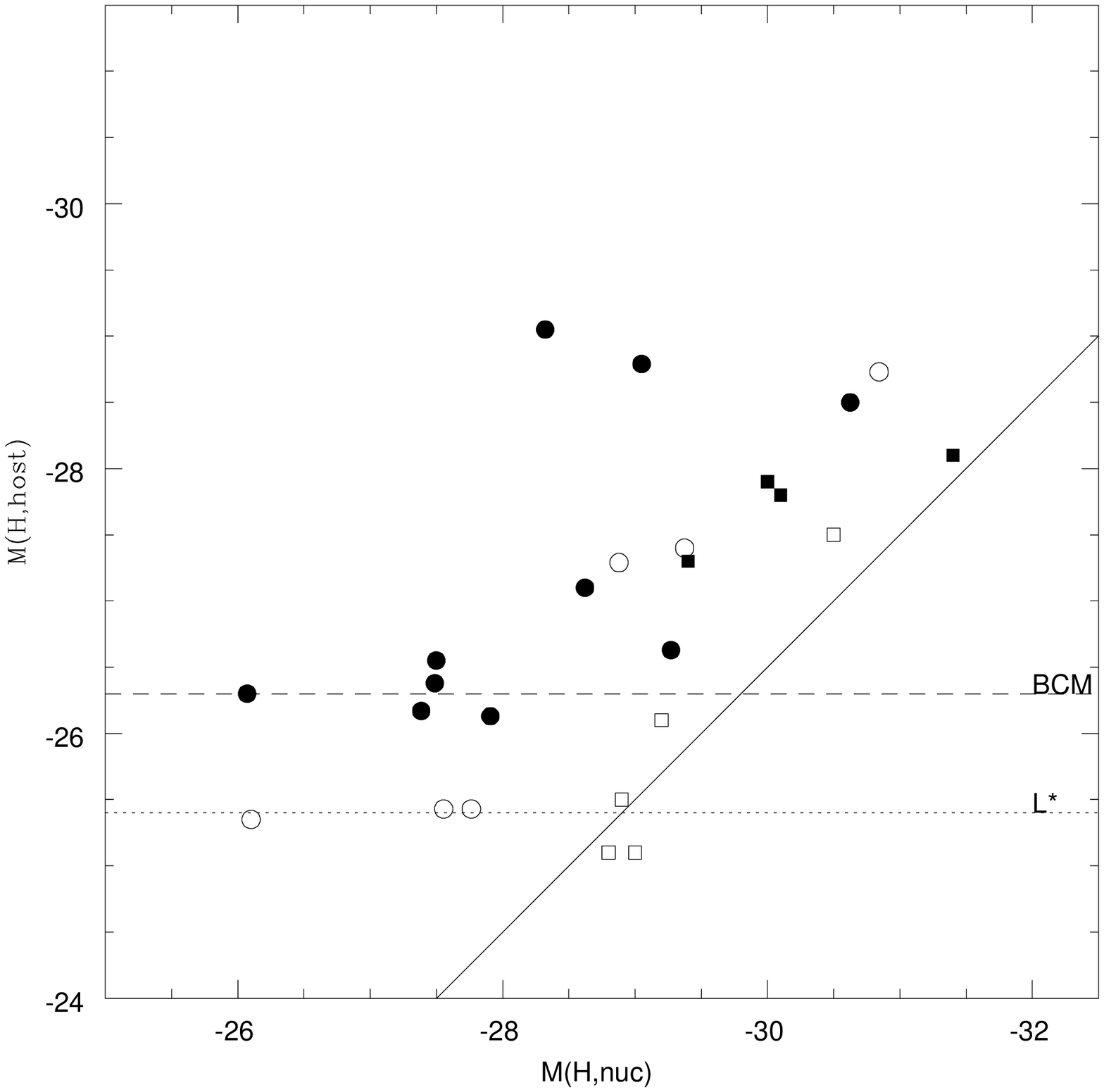,width=9cm,height=9cm}
\caption{The $H$-band nuclear vs. host luminosity. For symbols, see 
Fig.~\ref{fig:Mhz}. The solid line is the limiting mass--luminosity envelope 
from the M(B,nuc) vs. M(H,host) diagram of McLeod \& Rieke (1995), converted 
to $H$-band using a least squares fit of the M(B,nuc) and M(H,nuc) values for 
the SSRQs and FSRQs.}
\label{fig:MhMn}
\end{figure}

In Fig.~\ref{fig:MhMn}, we show the relation between the luminosities of the 
nucleus and the host galaxy for the SSRQs and FSRQs. We confirm for the SSRQs 
the tendency noted for FSRQs by KFS98 for the more powerful quasars to reside 
in more luminous hosts, in the sense that there appears to be a lower limit 
to the host luminosity, which increases with the quasar luminosity. All the 
SSRQs and FSRQs appear consistent with this boundary limit, first proposed by 
McLeod \& Rieke 1995). 

Recent dynamical investigations of nearby galaxies have reported evidence for 
supermassive black holes in their nuclei (Kormendy \& Richstone 1995; 
Magorrian et al. 1998). Assuming that quasar activity results from accretion 
of material onto a supermassive black hole, the nucleus-host relationship is 
in agreement with that found by Magorrian et al. (1998) from HST kinematic 
study between the mass (luminosity) of the black hole and the mass 
(luminosity) of the spheroidal component in nearby galaxies. Our results for 
SSRQs and FSRQs therefore suggest that the Magorrian et al. relationship 
extends to host galaxy masses at cosmological distances. On the other hand, 
the much weaker correlation found for more nearby, lower luminosity AGN (e.g. 
McLeod \& Rieke 1995; McLeod, Rieke \& Storrie-Lombardi 1999) may indicate 
that the onset of the correlation occurs only after a certain level in 
nuclear and/or galaxy luminosity has been reached. For example, at highest 
luminosities (highest black hole masses) quasars may accrete near the 
Eddington limit, triggering the onset of the host - black hole correlation 
(Lacy et al. 2000). 

However, we advise caution about obvious selection effects that may 
de-populate the upper left hand (faint nuclei are difficult to detect against 
luminous hosts) and lower right hand (faint hosts are difficult to detect 
under luminous nuclei) corners of the diagram. Moreover, it is not clear if 
this boundary limit depends on the radio loudness since a recent 
investigation of luminous RQQs at z $\sim$ 0.3 (Percival et al. 2000) does 
not find any correlation between host and nuclear luminosities.

\section{The close environment of SSRQs and FSRQs}

\begin{table}
\begin{center}
 \caption{The close environment of the FSRQs and SSRQs.\label{tab:env}}
 \begin{tabular}{llll}
\hline
Name         & z     & Nr(50 kpc) & Nr(100 kpc) \\
\hline
FSRQs:       &       &            &             \\
PKS 0336--019 & 0.852 & 0 & 1 \\
PKS 0403--123 & 0.571 & 0 & 3 \\
PKS 0405--123 & 0.574 & 0 & 2 \\
PKS 0420--014 & 0.915 & 0 & 0 \\
PKS 0440--003 & 0.844 & 1 & 1 \\
PKS 0454--463 & 0.858 & 1 & 1 \\
PKS 0605--085 & 0.872 & 0 & 1 \\
PKS 0637--752 & 0.654 & 1 & 1 \\
PKS 1055+018 & 0.888 & 0 & 3 \\
PKS 1253     & 0.538 & 0 & 0 \\
PKS 1504--166 & 0.876 & 0 & 0 \\
PKS 1954--388 & 0.626 & 0 & 3 \\
PKS 2128--123 & 0.501 & 0 & 1 \\
PKS 2145+067 & 0.990 & 0 & 0 \\
PKS 2243--123 & 0.630 & 0 & 1 \\
PKS 2345--167 & 0.576 & 0 & 1 \\
\hline
SSRQs:       &       &   &  \\
PKS 0056--00  & 0.717 & 0 & 0 \\
PKS 0159--11  & 0.669 & 0 & 0 \\
PKS 0349--14  & 0.614 & 0 & 0 \\
PKS 0413--21  & 0.807 & 0 & 1 \\
PKS 0414--06  & 0.773 & 0 & 0 \\
PKS 0454--22  & 0.534 & 0 & 0 \\
PKS 0518+16  & 0.759 & 0 & 0 \\
PKS 0710+11  & 0.768 & 1 & 1 \\
PKS 0825--20  & 0.822 & 0 & 1 \\
PKS 0838+13  & 0.684 & 0 & 0 \\
PKS 0855--19  & 0.660 & 0 & 1 \\
PKS 0903--57  & 0.695 & 0 & 0 \\
PKS 0959--44  & 0.937 & 0 & 2 \\
PKS 1046--40  & 0.620 & 0 & 0 \\
PKS 1116--46  & 0.713 & 0 & 2 \\
PKS 1136--13  & 0.554 & 0 & 0 \\
PKS 1237--10  & 0.750 & 0 & 0 \\
PKS 1327--21  & 0.528 & 0 & 1 \\
PKS 1335--06  & 0.625 & 0 & 1 \\
\hline
\end{tabular}
\end{center}
\end{table} 

It has been noted in many previous studies of quasar environments (e.g. 
Yee \& Green 1984; Stockton \& MacKenty 1987; Hutchings \& Neff 1990; 
Hutchings 1995; Hutchings et al. 1999) that quasars have often companions in 
their immediate environments and they are sometimes associated with disturbed 
morphology. Spectroscopy of these companions (e,g. Stockton 1978; Heckman et 
al. 1984) has shown that the companions are at the redshift of the quasar and 
are therefore physically associated. These observations have sustained the 
potentially important idea that nuclear activity can be triggered and/or 
fueled by strong tidal interactions and/or galaxy mergers (e.g. Heckman 1990)

Although both RLQs and RQQs seem to occur in dense groups of galaxies (and 
only rarely are located in rich galaxy clusters), a systematic difference of 
environment is found between RLQs and RQQs in the sense that the latter are 
generally found in poorer fields (e.g. Ellingson et al. 1991). There have 
also been some suggestions (Stockton 1978; Hutchings \& Neff 1990) that FSRQs 
(and in general, core-dominated RLQs) have a smaller frequency of companions 
with respect to the more common SSRQs. These differences, if confirmed, are 
important for understanding the connection between the characteristics of the 
nuclear activity and the environment.

To investigate these differences, we have compared the frequency of 
companions found around the matched samples of SSRQs and FSRQs. For both 
samples, we have counted the number of resolved companion objects (i.e. 
galaxies) within a specified radius and brighter than m$^*$(H) + 2, where 
m$^*$(H) is the apparent magnitude corresponding to M$^*$(H) = --25 at the 
redshift of each quasar. This limit is in all frames at least 1 mag brighter 
than the magnitude limit of the images. To evaluate the number of companion 
galaxies around the quasars, we have chosen two radii corresponding to 
projected distances of 50 and 100 kpc. At the average redshift of $\sim$0.7 
of the samples, 1 arcsec corresponds to $\sim$9.4 kpc. In Table~\ref{tab:env} 
we give for each quasar the number of resolved companions N(r).
 
The average number of companions around  FSRQs is 0.19$\pm$0.39 and 
1.19$\pm$1.01 within 50 and 100 kpc projected distance from the quasar, 
respectively. The corresponding values for the SSRQs are 0.05$\pm$0.22 and 
0.53$\pm$0.68. The fraction of FSRQs with no close companion is 81\% and 
25\% within 50 and 100 kpc from the quasar, respectively, while the 
corresponding fractions for the SSRQs are 95\% and 58\%. Contrary to previous 
suggestions (Stockton 1978; Hutchings \& Neff 1990), we find that the 
immediate environment of FSRQs is comparably rich (and possibly even richer) 
than that of SSRQs. This argues against a possible tight connection between 
the formation/evolution of the radio structure and the quasar galaxy 
environment. 

\section{Conclusions}

We have presented near--infrared $H$-band images of 19 SSRQs in the redshift 
range 0.5 $<$ z $<$ 1 with the aim of studying the quasar environment (host 
galaxies and companions) in comparison with our previously studied matched 
sample of FSRQs with respect to redshift and optical and radio luminosity. We 
are able to clearly detect the host galaxy in 10 (53 \%) SSRQs and marginally 
in 6 (32 \%) others, while the host remains unresolved in 3 (16 \%) cases. 
The galaxies hosting the SSRQs are large (average bulge scale-length 
R(e) = 9.0$\pm$1.7 kpc) and luminous (average M(H) = -27.2$\pm$1.1). They 
are, therefore, $\sim$2 mag more luminous than the typical galaxy luminosity 
L* (M*(H) = --25.0$\pm$0.2), and $\sim$1 mag more luminous than the brightest 
cluster galaxies (M(H) = --26.3$\pm$0.3). The SSRQ hosts appear to have 
similar luminosity as those of the FSRQ hosts (M(H)$\sim$--27). On the other 
hand, the average nucleus--to--galaxy luminosity ratio of SSRQs 
(LN/LG = 3.8$\pm$3.2) is significantly smaller than that found for the FSRQs 
(LN/LG = 21$\pm$11), in agreement with the idea that the latter are beamed. 
We find for SSRQs a positive trend, noted earlier for the FSRQs, between the 
host and nuclear luminosity. Finally, we find that the number of close 
companion galaxies is not strongly influenced by the characteristics of radio 
emission.  Comparable (and possibly larger in FSRQ) densities of companions 
are observed for FSRQs and SSRQs,

\begin{acknowledgements} 
This work was partly supported by the Italian Ministry for University and 
Research (MURST) under grants Cofin 98-02-32 and Cofin 98-02-15. This 
research has made use of the NASA/IPAC Extragalactic Database (NED), which is 
operated by the Jet Propulsion Laboratory, California Institute of 
Technology, under contract with the National Aeronautics and Space 
Administration.\\ 
\end{acknowledgements} 

\section*{Appendix: Notes on individual quasars and comparison with previous 
NIR photometry.}

\noindent{\bf PHL 923 = PKS 0056-00.} The host remains unresolved with upper 
limit M$_H$ $>$ -27.0, assuming R(e) = 10 kpc. 

\noindent{\bf 3C 57 = PKS 0159-11.} Hyland \& Allen (1982) measured H = 14.78 
in a 7 $''$ aperture, fainter than our H = 14.1 -- 14.2. The profile fit is 
rather poor with PSF only ($\chi^2$ = 3.7). Using a PSF + elliptical model 
results in a better fit ($\chi^2$ = 1.0), but there remains excess emission 
with respect to the first 2 pixels, indicating a bad match with the PSF. The 
host is thus only marginally resolved with R(e) = 8.8 kpc and M$_H$ = --25.4. 

\noindent{\bf 3C 95 = PKS 0349-14.} Neugebauer et al. (1979) measured 
H = 14.73 in a 7.5$''$ aperture, fainter than our H = 12.9. 

\noindent{\bf 3C 110 = PKS 0414-06.} Hyland \& Allen (1982) measured 
H = 14.71, while Sun \& Malkan (1989) measured H = 14.89, slightly fainter 
than our H = 14.6. This quasar was studied in the $R$-band by 
R\"onnback et al. (1996) who found it to be unresolved, with M(R) $>$ -20.9, 
whereas we resolved the host with M(H) = --26.6.

\noindent{\bf 3C 138 = PKS 0518+16.} The host galaxy is marginally resolved, 
with M$_H$ = -25.4 and R$_e$ = 8 kpc. This quasar was imaged with HST in the 
optical by De Vries et al. (1997), who claim it to be slightly extended 
(although they did not attempt any modelling).

\noindent{\bf 3C 175 = PKS 0710+11.} Simpson \& Rawlings (2000) measured 
H = 14.72$\pm$0.01 in a 3 arcsec aperture, fainter than our H = 13.1. The 
host is marginally resolved with M$_H$ = --28.7 and r$_e$ = 8 kpc. This 
quasar has not been resolved by previous ground-based optical studies 
(Malkan 1984; Hes, Barthel \& Fosbury 1996) but is marginally resolved with 
HST (Lehnert et al. 1999), showing roughly E-W orientation and a plume to 
S-SE up to 2 arcsec from the nucleus. 

\noindent{\bf PKS 0825-202.} The host remains unresolved with 
M(host) $>$ --25.0. 

\noindent{\bf 3C 207 = PKS 0838+13.} Simpson \& Rawlings (2000) measured 
H = 16.11$\pm$0.04 in a 3 arcsec aperture, in agreement with our H = 15.9. 
This quasar was not resolved by HST (Lehnert et al. 1999).

\noindent{\bf PKS 0903-57.} The profile fit is poor, resulting in a 
marginally resolved host with M$_H$ = -27.3 and R$_e$ = 4.3 kpc.

\noindent{\bf PKS 0959-443.} The profile fit is poor with PSF only. The 
quasar is marginally resolved with a reasonably good PSF + elliptical fit, 
yielding M$_H$ = -27.4 and R$_e$ = 11.1 kpc. Note that there is a bright star 
close to the quasar, and although it was removed prior to the fit, the 
possible detection of a host galaxy needs to be approached with caution.

\noindent{\bf PKS 1116-46.} This quasar was studied in the $R$-band by 
R\"onnback et al. (1996), who found it to be marginally resolved with 
M(R) = --23.0.

\noindent{\bf PKS 1136-138.} The rather short exposure time for this quasar 
resulted in limiting surface brightness of only $\mu_{lim}$ = 21. The quasar 
remains unresolved with a poor fit of the PSF in the center.

\noindent{\bf PKS 1327-21.} Glass (1981) measured K = 14.28$\pm$0.24 in a 
12$''$ aperture, brighter than our H = 15.1. Again, the fit with the PSF does 
not match well in the central pixel. The quasar is marginally resolved with a 
reasonably good PSF + elliptical fit, with M$_H$ = -25.4 and R$_e$ = 8.0 kpc. 
Wyckoff, Wehringer \& Gehren (1981) did not resolve the host in the optical 
with M(R) $>$ --21.9, whereas V\'eron-Cetty \& Woltjer (1990) claim the host to 
be resolved with M(V) = --23.7.

\noindent{\bf PKS 1335-06.} Romanishin \& Hintzen (1989) resolved the host 
galaxy in the optical with M(B) = -23.7 and R(e) = 7.2 kpc. The radio 
morphology of this quasar is a bent triple with bridges connecting the 
components (Price et al. 1993, Bogers et al. 1996). It appears that we have 
detected NIR emission from the NW hot spot and its associated radio lobe, to 
the NW and N of the quasar (Fig.~\ref{fig:cont}). However, the SE hot spot 
and lobe are not detected in our image.\\

\noindent{\bf References}\\

\noindent
Antonucci,R.R.J., 1993, ARA\&A 31, 473\\
Bahcall,J.N., Kirhakos,S., Saxe,D.H., Schneider,D.P., 1997, ApJ 479, 642\\
Blandford.R.D., Rees,M.J., 1978, Pittsburgh Conference on BL Lac Objects (ed. A.M.Wolfe), 328\\
Bogers,W.J., Hes,R., Barthel,P.D., Zensus,J.A., 1996, A\&AS 105, 91\\
Boyce,P.J., Disney,M.J., Blades,J.C. et al. 1998, MNRAS 298, 121\\
Bressan,A., Chiosi,C., Fagotto,F., 1994, ApJS 94, 63 \\
Capaccioli,M., Caon,N., D'Onofrio,M., 1992, MNRAS 259, 323\\
Carballo,R., Sanchez,S.F., Gonzalez-Serrano,J.L., Benn,C.R., Vigotti,M., 1998, AJ 115, 1234\\
de Vries,W.H., O'Dea,C.P., Baum,S.A. et al., 1997, ApJS 110, 191\\
Djorgovski,S., Davis,M., 1987, ApJ 313, 59\\
Eales,S., Rawlings,S., Law--Green,D., Cotter,G., Lacy,M., 1997, MNRAS 291, 593\\ 
Ellingson,E., Yee,H.K.C., Green,R.F., 1991, ApJ 371, 49\\
Fukugita,M., Shimasaku,K., Ichikawa,T., 1995, PASP 107, 945\\
Glass,I.S., 1981, MNRAS 194, 795\\
Heckman,T.M., 1990, in IAU Colloquim 124, p. 359\\
Heckman,T.M., Bothun,G.D., Balick,B., Smith,E.P., 1984 AJ 89, 958\\
Hes,R., Barthel,P.D., Fosbury,R.A.E., 1996, A\&A 313, 423\\
Hooper,E.J., Impey,C.D., Foltz,C.B., 1997, ApJ 480, L95\\
Hutchings,J.B., 1995, AJ 110, 994\\
Hutchings,J.B., Neff,S.G., 1990, AJ 99, 1715\\
Hutchings,J.B., Neff,S.G., 1992, AJ 104, 1\\
Hutchings,J.B., Crampton,D., Morris,S.L., Durand,D., Steinbring,E., 1999, AJ 117, 1109\\
Hyland,A.R., Allen,D.A., 1982, MNRAS 199, 943\\
Impey,C.D., Tapia,S., 1990, ApJ 354, 124\\
Kauffmann,G., H\"ahnelt,M., 2000, MNRAS 311, 576\\
Kormendy,J., 1977, ApJ 218, 333\\
Kormendy,J., Richstone,D., 1995, ARA\&A 33, 581\\
Kotilainen,J.K., Falomo,R., Scarpa,R., 1998, A\&A 332, 503 (KFS98)\\
Lacy,M., Bunker,A.C., Ridgway,S.E., 2000, AJ 120, 68\\
Landolt,A., 1992, AJ 104, 340\\
Ledlow,M.J., Owen,F.N., 1996, AJ 112, 9\\
Lehnert,M.D., Heckman,T.M., Chambers,K.C., Miley,G.K., 1992, ApJ 393, 68\\
Lehnert,M.D., Miley,G.K., Sparks,W.B., et al., 1999, ApJS 123, 351\\
Lidman,C., Cuby,J.G., 1998, SOFI User Manual, ESO\\
Lilly,S.J., Longair,M.S., 1984, MNRAS 211, 833\\
Lowenthal,J.D., Heckman,T.M., Lehnert,M.D., Elias,J.H., 1995, ApJ 439, 588\\
Magorrian,J., Tremaine,S., Richstone,D. et al., 1998, AJ 115, 2285\\
Malkan,M.A., 1984, ApJ 287, 555\\
McLeod,K.K., Rieke,G.H., 1994a, ApJ 420, 58\\
McLeod,K.K., Rieke,G.H., 1994b, ApJ 431, 137\\
McLeod,K.K., Rieke,G.H., 1995, ApJ 454, L77\\
McLeod,K.K., Rieke,G.H., Storrie-Lombardi,L.J., 1999, ApJ 511, L67\\
McLure,R.J., Dunlop,J.S., 2000, MNRAS, in press (astro-ph/9908214)\\
McLure,R.J., Kukula,M.J., Dunlop,J.S., et al., 1999, MNRAS 308, 377\\
Mobasher,B., Sharples,R.M., Ellis,R.S., 1993, MNRAS 263, 560\\
Moffat,A.F.J., A\&A 3, 455\\
Moorwood,A.F.M. et al., 1992, ESO Messenger 69, 61\\
Neugebauer,G., Oke,J.B., Becklin,E.E., Matthews,K., 1979, ApJ 230, 79\\
Padovani,P., Urry,C.M., 1992, ApJ 387, 449\\
Percival,W.J., Miller,L., McLure,R.J., Dunlop,J.S., 2000, MNRAS, in press (astro-ph/0002199)\\
Poggianti,B.M., 1997, A\&AS 122, 399\\
Price,R., Gower,A.C., Hutchings,J.B. et al., 1993, ApJS 86, 365\\
Quirrenbach,A. Witzel,A., Krichbaum,T.P. et al., 1992, A\&A 258, 279\\
Romanishin,W., Hintzen,P., 1989, ApJ 341, 41\\
R\"onnback,J., van Groningen,E., Wanders,I., \"{O}rndahl,E., 1996, MNRAS 283, 282\\
Silk,J., Rees,M.J., 1998, A\&A 331, L1\\
Simpson,C., Rawlings,S., 2000, MNRAS, in press (astro-ph/0005570)\\
Small,T.A., Blandford,R.D., 1992, MNRAS 259, 725\\
Smith,E.P., Heckman,T.M., 1989, ApJ 341, 658\\
Spinrad,H., Dey,A., Stern,D., et al., 1997, ApJ 484, 581\\
Stiavelli,M., Treu,T., Carollo,C.M. et al., 1999, A\&A 343, L25\\
Stockton,A., 1978, ApJ 223, 747\\
Stockton,A., MacKenty,J.W., 1987, ApJ 316, 584\\
Sun,W.H., Malkan,M.A., 1989, ApJ 346, 68\\
Taylor,G.L., Dunlop,J.S., Hughes,D.H., Robson,E.I., 1996, MNRAS 283, 930\\
Thuan,T.X., Puschell,J.J., 1989, ApJ 346, 34\\
Urry,C.M., Padovani,P., 1995, PASP 107, 803\\
Vermeulen,R.C., Cohen,M.H., 1994, ApJ 430, 467\\
V\'eron--Cetty,M--.P., Woltjer,L., 1990, A\&A 236, 69\\
V\'eron--Cetty,M--.P., V\'eron,P., 1998, ESO Scientific Report\\
Wyckoff,S., Wehinger,P.A., Gehren,T., 1981, ApJ 247, 750\\
Yee,H.K.C., Green,R.F., 1984, ApJ 280, 79\\
 
\end{document}